\RequirePackage{lineno}
\documentclass[prd,showpacs,twocolumn,groupedaddress,superscriptaddress]{revtex4}
\usepackage{graphicx}  % needed for figures

\usepackage{dcolumn}   % needed for some tables

\usepackage{bm}        % for math

\usepackage{amssymb}   % for math
\usepackage{amsmath}
\usepackage{setspace}   % for dspace

\usepackage{ctable}   % JMP: To use footnotes with tables....
\usepackage{multirow}  % JMP: Make tables that can have entries that span several rows
\usepackage{subfigure} % JMP: Divide figures into subfigures
\usepackage{lineno}

\newcommand{\dzero}     {D0}

\newcommand{\sherpa}    {{\sc sherpa}}

\newcommand{\met}       {\mbox{$\not\!\!E_T$}}

\newcommand{\pt}        {$p_T$}

\begin{document}

% the following information is for internal review, please remove them for submission
\widetext
%\rightline{by Dec. 06, 2004}

\hspace{5.2in}\mbox{FERMILAB-PUB-11-400-E}

\title{Model independent search for new phenomena in $\boldsymbol{p \bar{p}}$ collisions at $\bm{\sqrt{s}=}$1.96 TeV}

%\input list_of_authors_r2.tex  % input Dzero author list
% remove these 3 lines before journal submittal.
%\centerline{author list dated 2 August 2011}
% end removal before journal submittal
%
\affiliation{Universidad de Buenos Aires, Buenos Aires, Argentina}
\affiliation{LAFEX, Centro Brasileiro de Pesquisas F{\'\i}sicas, Rio de Janeiro, Brazil}
\affiliation{Universidade do Estado do Rio de Janeiro, Rio de Janeiro, Brazil}
\affiliation{Universidade Federal do ABC, Santo Andr\'e, Brazil}
\affiliation{Instituto de F\'{\i}sica Te\'orica, Universidade Estadual Paulista, S\~ao Paulo, Brazil}
\affiliation{University of Science and Technology of China, Hefei, People's Republic of China}
\affiliation{Universidad de los Andes, Bogot\'{a}, Colombia}
\affiliation{Charles University, Faculty of Mathematics and Physics, Center for Particle Physics, Prague, Czech Republic}
\affiliation{Czech Technical University in Prague, Prague, Czech Republic}
\affiliation{Center for Particle Physics, Institute of Physics, Academy of Sciences of the Czech Republic, Prague, Czech Republic}
\affiliation{Universidad San Francisco de Quito, Quito, Ecuador}
\affiliation{LPC, Universit\'e Blaise Pascal, CNRS/IN2P3, Clermont, France}
\affiliation{LPSC, Universit\'e Joseph Fourier Grenoble 1, CNRS/IN2P3, Institut National Polytechnique de Grenoble, Grenoble, France}
\affiliation{CPPM, Aix-Marseille Universit\'e, CNRS/IN2P3, Marseille, France}
\affiliation{LAL, Universit\'e Paris-Sud, CNRS/IN2P3, Orsay, France}
\affiliation{LPNHE, Universit\'es Paris VI and VII, CNRS/IN2P3, Paris, France}
\affiliation{CEA, Irfu, SPP, Saclay, France}
\affiliation{IPHC, Universit\'e de Strasbourg, CNRS/IN2P3, Strasbourg, France}
\affiliation{IPNL, Universit\'e Lyon 1, CNRS/IN2P3, Villeurbanne, France and Universit\'e de Lyon, Lyon, France}
\affiliation{III. Physikalisches Institut A, RWTH Aachen University, Aachen, Germany}
\affiliation{Physikalisches Institut, Universit{\"a}t Freiburg, Freiburg, Germany}
\affiliation{II. Physikalisches Institut, Georg-August-Universit{\"a}t G\"ottingen, G\"ottingen, Germany}
\affiliation{Institut f{\"u}r Physik, Universit{\"a}t Mainz, Mainz, Germany}
\affiliation{Ludwig-Maximilians-Universit{\"a}t M{\"u}nchen, M{\"u}nchen, Germany}
\affiliation{Fachbereich Physik, Bergische Universit{\"a}t Wuppertal, Wuppertal, Germany}
\affiliation{Panjab University, Chandigarh, India}
\affiliation{Delhi University, Delhi, India}
\affiliation{Tata Institute of Fundamental Research, Mumbai, India}
\affiliation{University College Dublin, Dublin, Ireland}
\affiliation{Korea Detector Laboratory, Korea University, Seoul, Korea}
\affiliation{CINVESTAV, Mexico City, Mexico}
\affiliation{Nikhef, Science Park, Amsterdam, the Netherlands}
\affiliation{Radboud University Nijmegen, Nijmegen, the Netherlands and Nikhef, Science Park, Amsterdam, the Netherlands}
\affiliation{Joint Institute for Nuclear Research, Dubna, Russia}
\affiliation{Institute for Theoretical and Experimental Physics, Moscow, Russia}
\affiliation{Moscow State University, Moscow, Russia}
\affiliation{Institute for High Energy Physics, Protvino, Russia}
\affiliation{Petersburg Nuclear Physics Institute, St. Petersburg, Russia}
\affiliation{Instituci\'{o} Catalana de Recerca i Estudis Avan\c{c}ats (ICREA) and Institut de F\'{i}sica d'Altes Energies (IFAE), Barcelona, Spain}
\affiliation{Stockholm University, Stockholm and Uppsala University, Uppsala, Sweden}
\affiliation{Lancaster University, Lancaster LA1 4YB, United Kingdom}
\affiliation{Imperial College London, London SW7 2AZ, United Kingdom}
\affiliation{The University of Manchester, Manchester M13 9PL, United Kingdom}
\affiliation{University of Arizona, Tucson, Arizona 85721, USA}
\affiliation{University of California Riverside, Riverside, California 92521, USA}
\affiliation{Florida State University, Tallahassee, Florida 32306, USA}
\affiliation{Fermi National Accelerator Laboratory, Batavia, Illinois 60510, USA}
\affiliation{University of Illinois at Chicago, Chicago, Illinois 60607, USA}
\affiliation{Northern Illinois University, DeKalb, Illinois 60115, USA}
\affiliation{Northwestern University, Evanston, Illinois 60208, USA}
\affiliation{Indiana University, Bloomington, Indiana 47405, USA}
\affiliation{Purdue University Calumet, Hammond, Indiana 46323, USA}
\affiliation{University of Notre Dame, Notre Dame, Indiana 46556, USA}
\affiliation{Iowa State University, Ames, Iowa 50011, USA}
\affiliation{University of Kansas, Lawrence, Kansas 66045, USA}
\affiliation{Kansas State University, Manhattan, Kansas 66506, USA}
\affiliation{Louisiana Tech University, Ruston, Louisiana 71272, USA}
\affiliation{Boston University, Boston, Massachusetts 02215, USA}
\affiliation{Northeastern University, Boston, Massachusetts 02115, USA}
\affiliation{University of Michigan, Ann Arbor, Michigan 48109, USA}
\affiliation{Michigan State University, East Lansing, Michigan 48824, USA}
\affiliation{University of Mississippi, University, Mississippi 38677, USA}
\affiliation{University of Nebraska, Lincoln, Nebraska 68588, USA}
\affiliation{Rutgers University, Piscataway, New Jersey 08855, USA}
\affiliation{Princeton University, Princeton, New Jersey 08544, USA}
\affiliation{State University of New York, Buffalo, New York 14260, USA}
\affiliation{Columbia University, New York, New York 10027, USA}
\affiliation{University of Rochester, Rochester, New York 14627, USA}
\affiliation{State University of New York, Stony Brook, New York 11794, USA}
\affiliation{Brookhaven National Laboratory, Upton, New York 11973, USA}
\affiliation{Langston University, Langston, Oklahoma 73050, USA}
\affiliation{University of Oklahoma, Norman, Oklahoma 73019, USA}
\affiliation{Oklahoma State University, Stillwater, Oklahoma 74078, USA}
\affiliation{Brown University, Providence, Rhode Island 02912, USA}
\affiliation{University of Texas, Arlington, Texas 76019, USA}
\affiliation{Southern Methodist University, Dallas, Texas 75275, USA}
\affiliation{Rice University, Houston, Texas 77005, USA}
\affiliation{University of Virginia, Charlottesville, Virginia 22901, USA}
\affiliation{University of Washington, Seattle, Washington 98195, USA}
\author{V.M.~Abazov} \affiliation{Joint Institute for Nuclear Research, Dubna, Russia}
\author{B.~Abbott} \affiliation{University of Oklahoma, Norman, Oklahoma 73019, USA}
\author{B.S.~Acharya} \affiliation{Tata Institute of Fundamental Research, Mumbai, India}
\author{M.~Adams} \affiliation{University of Illinois at Chicago, Chicago, Illinois 60607, USA}
\author{T.~Adams} \affiliation{Florida State University, Tallahassee, Florida 32306, USA}
\author{G.D.~Alexeev} \affiliation{Joint Institute for Nuclear Research, Dubna, Russia}
\author{G.~Alkhazov} \affiliation{Petersburg Nuclear Physics Institute, St. Petersburg, Russia}
\author{A.~Alton$^{a}$} \affiliation{University of Michigan, Ann Arbor, Michigan 48109, USA}
\author{G.~Alverson} \affiliation{Northeastern University, Boston, Massachusetts 02115, USA}
\author{G.A.~Alves} \affiliation{LAFEX, Centro Brasileiro de Pesquisas F{\'\i}sicas, Rio de Janeiro, Brazil}
% ...
%
% visitor_addresses.tex                        2 August 2011
%  available symbols are:
%  $\ast, \dag, \ddag, \S, \P, $\|$, $\ast\ast$, \dag\dag, \ddag\ddag ,\#
%
\collaboration{The D0 Collaboration\footnote{with visitors from
%{alton}
$^{a}$Augustana College, Sioux Falls, SD, USA,
%{burdin}
$^{b}$The University of Liverpool, Liverpool, UK,
%{falkowski}
%$^{?}$Laboratoire de Physique Theorique, Orsay, FR
%{garcia-guerra}
$^{c}$UPIITA-IPN, Mexico City, Mexico,
%{haas,partridge}
$^{c}$SLAC, Menlo Park, CA, USA,
%{hesketh}
$^{e}$University College London, London, UK,
%{luna-garcia}
$^{f}$Centro de Investigacion en Computacion - IPN, Mexico City, Mexico,
%{podesta-lerma}
$^{g}$ECFM, Universidad Autonoma de Sinaloa, Culiac\'an, Mexico,
and 
%{weber}
$^{h}$Universit{\"a}t Bern, Bern, Switzerland.
%{hooper}
%$^{?}$Visitor from Bradley University, Peoria, IL, USA.
%{kozminski}
%$^{?}$}Visitor from Lewis University, Romeoville, IL, USA.
%{deceased}
$^{\ddag}$Deceased.
}} \noaffiliation
\vskip 0.25cm

\date{August 26, 2011}

\begin{abstract}

We describe a model independent search for physics 
beyond the standard model in lepton final states.  We examine 117 
final states using 1.1 fb$^{-1}$ of $p \bar{p}$ collisions data at $\sqrt{s} = 1.96$ TeV 
collected with the D0 detector.  We conclude that all observed discrepancies between data and model can be 
attributed to uncertainties in the standard model background modeling, and hence we do not see any evidence 
for physics beyond the standard model.

\end{abstract}


\maketitle

\section{\label{sec:intro}Introduction}
The standard model (SM)
%has been remarkably successful in accommodating all the fundamental particles, 
%with the notable exception of the Higgs boson \cite{chengli}.
has been remarkably successful in accommodating all the interactions between the fundamental particles \cite{chengli}.
Despite this success, there are strong motivations 
to expect new phenomena at energies at the order of the electroweak
scale. For example, the Higgs boson \cite{guralnik} receives quantum corrections
to its mass through loop diagrams. The scalar nature of the Higgs
boson leads to a quadratic divergence, with an upper limit of the
integral set by the highest scale, i.e., the Planck mass
($10^{19}$ GeV). To maintain the Higgs mass close to the
electroweak scale, it is necessary to fine tune a parameter in the
theory to within $M_W/M_{\text{Planck}} \approx 10^{-16}$ \cite{hierarchy}.

There are few logical options for overcoming this problem. If the
Higgs boson does not exist, then there must be a new contribution to the physics at the
electroweak scale.  If the Higgs boson does exist, then the theory must be either
 fine tuned or a generalized Higgs scheme, beyond the SM, is present at the electroweak scale.  

%However, if the Higgs boson exists and the
%theory is fine tuned, one can avoid the need for new physics at
%the electroweak scale.

Assuming that beyond standard model (BSM) physics exists, we do not know how it appears, rendering its search difficult.
While there are many theories that predict observable differences with the SM, these
models usually depend on additional unspecified parameters which broaden the possible range of results.

%There are many theories that predict observable differences with the SM, but they generally do not specify 
%precise parameters for a search. 
Motivated by uncertainty and expectations of physics beyond the SM, we examined data from many channels in 
$p \bar{p}$ collisions at $\sqrt{s} = $ 1.96 TeV at the Tevatron Collider at 
Fermilab, collected by the D0 experiment, for deviations from the SM. 
After this, we focus on events with objects 
with high transverse momentum ($p_T$) in a quasi-model-independent search of new phenomena effects. 
Our background model is specific for final states
containing leptons, which form the focus of this paper. Similar approaches have
been applied to data from the D0 Collaboration ~\cite{run1sleuth1,run1sleuth2,run1sleuth3},
the H1 Collaboration at the HERA $ep$ collider at DESY~\cite{heraMIS}, and the CDF 
Collaboration at the Tevatron~\cite{cdfPRD,cdfRC}.

Our technique trades the sensitivity of specific searches for
breadth of coverage:  we do not design selections focused on a
particular model and neglect systematic uncertainties. This way, we can incorporate many
channels without developing a detailed modeling for each individual channel.  
This approach limits sensitivity for physics beyond the SM in individual final states, but it helps identify global differences relative to the SM expectations.
%to unique features of individual final states it helps point to global 
%differences relative to the SM expectations.  
If any particular final state or distribution found discrepant with the SM remains significantly 
discrepant after systematic uncertainties are considered, then it warrants claim for the presence of physics beyond the SM. 
The benefit of this approach is that we can look in a coordinated way at many channels,
applying expectations from the SM and a
model of the detector in a relatively straightforward manner, to search for discrepancies between data and the SM.  
%We have also chosen to
%focus on high transverse momentum ($p_T$) phenomena.

%The search for new physics starts with selecting only events with objects that have large $p_T$ values collected by the \dzer\ experiment.
% To enter our data set, high thresholds are set on
%objects and events are saved in a reduced format. The net effect of the format and the high-$p_T$
%selection requires a factor of $10^4$ less storage volume than the
%standard full \dzer\ data set. The motivation behind this compact data set is
%that a model independent search requires frequent cycling over very large data
%and Monte Carlo (MC) samples and running over the full \dzer\ data set would
%significantly slow down the analysis.

%The data for the search consists of events containing objects that have only
%large $p_T$ values.  This selection, and saving the chosen events with a reduced format, provides a reduction factor of 
%$10^4$ in storage volume.

%Our Monte Carlo does not properly 
%model background that arises from purely Quantum Chromodynamics (QCD) processes, such as jets 
%faking electrons.  To account for this, we reverse some of our 
%object selection cuts in data to produce samples with electron, 
%muon, or tau objects which are mostly fakes from QCD.

The data for the search consists of events containing objects that have large $p_T$ values.  
We divide the data and the selected Monte Carlo (MC) simulated events into seven inclusive subsets based on the number and types of leptons 
identified in each event.  Unlike the search conducted by the CDF Collaboration~\cite{cdfPRD,cdfRC}, only events with at 
least one electron or muon are considered.  For each of the chosen 
final states, we apply corrections to the MC simulation, as determined from the previous D0 studies based on well-understood regions of phase space, 
dominated by particular SM processes, as discussed in Sec.~\ref{sec:evgenpre}. 
To account for any incorrect normalizations in the absence of 
systematic uncertainties, we fit for contributions from each of the subsets 
to obtain scale factors which reproduce the distributions in the 
selected data with MC events and multijet background events determined from data, as discussed in Sec.~\ref{sec:incstates}.

The seven non-overlapping inclusive subsets are merged to create 
an input file for the analyses employing algorithms called {\sc vista} and { \sc sleuth} ~\cite{cdfPRD}, as discussed in Sec.~\ref{sec:exstates}.

%outline strategy (no details, handled in appropriate sections).

\section{\label{sec:d0det}\dzero~ Detector}
The data correspond to $1.07 \pm 0.07$ fb$^{-1}$ of integrated luminosity from $p\overline{p}$ collisions at the Tevatron 
Collider at Fermilab, collected with the \dzero\ detector at $\sqrt{s}=1.96$ TeV during 2002--2006.

 The \dzero\ detector is described in detail elsewhere~\cite{run2det}.
The central tracking, calorimetry, and muon systems are the components
most important to this analysis.
The central tracking system consists of a
silicon microstrip tracker (SMT) and a central fiber tracker (CFT),
both located within a 2 T superconducting solenoidal
magnet, and provides charged particle tracking for pseudorapidities $|\eta|<3$,
where $\eta = - \ln [\tan(\theta/2)]$, and $\theta$ is the polar angle relative to the center of the detector with respect to the proton beam direction.

The three liquid-argon/uranium calorimeters are
housed in separate cryostats.  Outside of the tracking system, a central section covers up to $|\eta|=1.1$. Two end calorimeters
extend coverage to $|\eta|=4.2$.    The calorimeter is
highly segmented with four electromagnetic (EM) and four to five hadronic
longitudinal layers; transverse to the particle direction, typical segmentation is $\Delta \eta = \Delta \phi = 0.1$, where $\phi$
is the azimuthal angle.

Beyond the calorimeter, a muon system consists of a layer of tracking detectors 
and scintillation trigger counters in front of 1.8 T iron toroids, 
followed by two similar layers after the toroids, all at pseudorapidities 
$|\eta|<2.0$~\cite{run2muon}.

A three level trigger system selects events, recording data at about 100 Hz.  Our sample was collected using triggers that 
select events with at least one electron or one muon.

%We apply data selections to ensure that
%events with known detector problems are excluded from the analysis.
%The nature of our analysis is such that we are not concerned about 
%absolute trigger efficiencies, which are accounted for in our 
%scale factors which are described later in this paper. Also,
%the $p_T$ dependence of the trigger efficiency at low $p_T$ 
%does not affect the shapes of our distributions due to high $p_T$ cuts.

%Boilerplate summary, refer to paper for details

\section{\label{sec:object_id}Object ID and Event Selection}
%\section{Object Selection \label{sec:eventselection}}
In this section, we describe the 
identification criteria used to select energetic
objects isolated from other event activity, {\it viz.,} electrons ($ e^\pm$),
muons ($ \mu^\pm$), tau leptons ($\tau^{\pm}$), missing transverse energy (\met), jets, and $b$-quark jets. The selection criteria for all these
objects are identical for all final states.
%In addition we also define the criteria for electron/muon
%which can mimic the signatures of the said isolated leptons
%-- they are called non-isolated electrons/muons.  
In addition, we discuss the criteria for non-isolated electrons and muons, i.e.,
objects that are not truly isolated but can mimic the signatures of isolated leptons. Because of the difficulty of modeling such false leptons,
their contributions are estimated directly from data.

\subsection{Vertices}
Only $p\bar{p}$ interaction vertices reconstructed from at least three tracks are allowed in this analysis.  
Based on the $p_T$ of the tracks associated with that vertex, we define the primary $p\bar{p}$ interaction vertex (PV), as the one with smallest probability of originating  
from a  minimum-bias interaction \cite{PV}. 
The $z$ coordinate of the PV ($z_{\text{PV}}$) is required to be $|z_{\text{PV}}|<$ 60 cm (where the $z$ axis is the
axis along beam direction, with origin at the center of the detector).
%The PV was required to be within 60 cm of the
%center of the detector along the z axis.

\subsection{Electrons and Photons}
Electrons are characterized by an isolated shower in the calorimeter and an
isolated track in the central tracker. Starting with a seed cell, a calorimeter cluster is formed 
using cells within a cone of radius $\rm \Delta {\cal R} <0.4$ where $\rm \Delta {\cal R} = 
\sqrt{(\Delta\eta)^2+(\Delta\phi)^2}$.
Such clusters are required to pass the calorimeter isolation criterion 
$(E_{\rm{tot}}( \Delta {\cal R}  < 0.4) - E_{\rm{EM}}(  \Delta {\cal R}  < 0.2))/E_{\rm{EM}}( \Delta {\cal 
R}  < 0.2) < 0.2$, where $E_{\rm{tot}}$ 
is the total energy of the shower, summing the EM and hadronic calorimeter cells, and $E_{\rm{EM}}$ is the energy in the EM calorimeter only. 
Every accepted cluster must have 90\% of $E_{\rm{tot}}$ within the EM calorimeter, 
pass a $\chi^{2}$-based selection on the spatial distribution of the shower, and be matched with a track extrapolated from the central tracker.
%An electron likelihood cut, based on seven tracking and calorimeter parameters, is used to further enhance electron purity. 
An electron likelihood ($L_e$), based on seven tracking and calorimetric parameters, is used to enhance 
signal purity of the candidate electrons. 
Photons are identified as electromagnetic clusters that pass the same isolation and shower criteria,
 but fail to match with a track. 

%With different sets of cuts on the aforesaid electron
%variables, D\O\  has several definitions for electrons and they are unified
%together in a package, emid\_cuts. 
In this analysis, we use only electrons that are found in the central calorimeter (CC), 
with $|\eta|<1.1$ and $p_T > 15$ GeV.  Typical electron detection efficiencies are 70\% to 80\%.

To estimate the contribution from non-isolated electrons (e.g. from multijet background), we use the same selection as for signal, but with a 
reversed $L_e$ likelihood criterion.

%Photons are reconstructed in a similar manner, except that they are required to have an 
%EM fraction of 95\%, have no likelihood cut, and must not have a matched track.

%In this analysis, we classify the electrons into
%two categories depending on their positions in the calorimeter: CC ($\rm |\eta|<1.1$)
%and EC ($\rm 1.5<|\eta|<2.5$) having $\rm E_T>35$ GeV. 
%and use the ``Tight\_hmx\_trk''
%definition from p18-br-03 version of emid\_cuts. 

\subsection{Muons}
Muons are identified in the muon system, and then
matched to tracks.  They are required to have $|\eta|<1.5$ and $p_T>15$ GeV. The track requirements include a selection on
DCA $<$ 0.02 (0.2) cm 
for tracks with (without) hits in the SMT, where DCA is the distance of closest approach of the track to the PV in the transverse 
plane.

%The muons produce a signal while passing through the calorimeter. 
%Therefore, isolated muon trajectory would be characterized
%by least activity in the calorimeter around the muon track and tracker isolation
%would further enhance the isolated muon purity. 
We require muons to be isolated, meaning that the sum of the transverse energies in calorimeter cells in an annular region ($\rm 
0.1<\Delta {\cal R} <0.4$) around the muon track, and the sum of the tracks $p_T$ in a cone of $\rm \Delta {\cal R} <0.5$ around 
the muon track must both be less than 2.5 GeV.

To estimate the multijet background in the single muon sample, we use control samples where the isolation variables are required 
to be between 2.5 GeV and 8 GeV.  All other criteria are the same as in the signal data sample.

Because the muon $p_T$ is estimated by the $p_T$ of the matching track in the central tracker, the momentum resolution decreases with 
increasing $p_T$.  To restrict the analysis to muons with well measured momenta, we require the significance of its $p_T$ measurement to be 
($1/p_{T})/\sigma(1/p_{T}) > 3$, where $\sigma(1/p_{T})$ is the uncertainty on the measurement of the track curvature (inverse of the muon 
track's $p_T$).  This effectively limits muons to $p_T < 200$ GeV.

%To estimate QCD backgrounds due to non-isolated muons, we require the isolation variables to be between 2.5 GeV and 8 GeV.

%Kinematic requirements include $\rm |\eta|<1.5$ and $p_T>20$ GeV.
%Similarly the ``TrackHalo'' is defined as the sum of 

%In this analysis we use ``MEDIUM'', Nseg3 muons having $\rm p_T>20$ GeV and $\rm |\eta|<2.0$. 
%The muons are required
%to have ``track\_medium''~\cite{d05157} quality match with the tracks in the
%central tracker. 

%For isolated muons we require both TrackHalo and CalorimeterHalo
%to be less than 2.5 GeV. 
%However for the fake/non-isolated muons, we select ``LOOSE''
%nseg2 muons having the values of ``TrackHalo'' and ``CalorimeterHalo'' in the
%range of 2.5-8 GeV.

\subsection{Tau Leptons} 
Tau leptons can decay to $e\nu_e\nu_{\tau}$, $\mu\nu_{\mu}\nu_{\tau}$, or hadrons $h\nu_{\tau}$ ($\tau_h$). 
It is difficult to determine whether a light lepton 
in an event originated from a $\tau$, but the signature from $\tau_h\to h\nu_{\tau}$ differs significantly from that of a jet. 
The decays $\tau\to\pi\nu_{\tau}$ are referred to as Type-1. Decays corresponding to $\tau^{\pm}\to\pi^{\pm}n\pi^0\nu_{\tau}$ 
are referred to as Type-2 ($n$ is an integer $\geq 1$), and decays to multiple charged
pions are referred to as Type-3 decays.  Type-3 decays differ
from Type-1 ($\tau_1$) and Type-2 ($\tau_2$) by being matched to multiple tracks, and are not used
in this analysis.
Type-1 and Type-2 decays are required to have
$|\eta|<1.1$ and a track with at least one SMT hit, as well as $p_T>10$ GeV for Type-1, and 
$p_T>5$ GeV for Type-2 tau leptons.  
%For $\tau$-type 3 the track 
%requirements are that $\sum p_T>10$ GeV and the sum of the track charges be $\not=0$ 
%($\tau$-type 3 can have 2 or 3 tracks). 
There are also requirements concerning overlaps of objects:
%$\rm \Delta R(\mu,\tau)>0.4$ where $\mu$ can be ``LOOSE'', nseg2 with no isolation requirement
$\rm \Delta {\cal R} (\mu,\tau)>0.4$ and ${\rm \Delta {\cal R}}(e,\tau)>0.4$, where $\tau$, $\mu$ and $e$ are as defined 
above, except that muons that pass the overlap criterion do not have to pass the additional isolation requirement.  To distinguish 
$\tau_h$ decays from jets, we use a neural network discriminant~\cite{tauneuralnet}, $\rm NN_h$, and to 
distinguish 
Type-2 $\tau_{h}$ from electrons, we use an additional neural network, $\rm NN_e$.  We require 
%<<<<<<< objectid.tex
%that $NN_h>0.9$ for type-1 and -2 taus and $NN_e>0.2$ for type-2 taus. 
%For $e+\tau$ events there is also the requirement that 
%$\tau$-type 2 candidates not be in an EM calorimeter $\phi$ crack.
%=======
$\rm NN_h>0.9$ for $\tau_1$ and $\tau_2$, and $\rm NN_e>0.2$ for $\tau_2$. 
%For $e+\tau$ events there is also the requirement that type-2 tau-candidates not be in an EM calorimeter $\phi$ crack.

%>>>>>>> 1.20

To model the multijet contribution to final states with $\tau_h$ decays, we select events with $\tau_h$ candidates as above, but with
$0.3 < {\rm NN_h} < 0.8$.
% and do not invert the likelihood cut for electrons or the isolation cut for muons.

\subsection{Jets \label{calorimeterjetsection}}
We reconstruct jets within $|\eta|<2.5$, using an iterative midpoint cone algorithm \cite{jetcone}
with cone radius of 0.5 and a minimum $p_T$
requirement of 20 GeV after applying a jet energy scale (JES) correction as discussed in Sec.~\ref{jes_corr}. 
%In order to eliminate fake jets
%(from calorimeter noise) and to separate them from electrons/photons,
%the jets are also required to pass several Jet ID criteria~\cite{jetid-note}.
%In addition overlaps with taus ($\rm \Delta {\cal R} <0.4$) are removed from the jet list.
%Jets within a cone with a radius of $\rm \Delta {\cal R} < 0.1$ from $\tau_h$ or electrons are removed from consideration.
Jets separated from a $\tau_h$ or an electron by $\rm \Delta {\cal R} < 0.5$ are removed from consideration.

\subsection{b-jets \label{btaggingdetails}}
Bottom and charm quarks can travel measurable distances from the PV before decaying, so that their decay products 
originate from an identifiable secondary vertex.  
This provides a way of tagging jets coming from a $b$($c$)-quark decay by examining the 
associated tracks~\cite{btag}.  Before applying any $b$-tagging criteria, the jets are required to pass
both calorimeter criteria outlined in Sec.~\ref{calorimeterjetsection} and the taggability criteria.  
A jet is taggable if it is matched to a track jet,
which is a jet formed from tracks, reconstructed using a
simple cone-clustering algorithm of $\Delta {\cal R} < 0.5$. At least two
tracks are required, with at least one having $p_T > 1$ GeV and 
another with $p_{T} > 0.5$ GeV. Every track in the jet is required to have at
least one hit in the SMT detector, a DCA $< 0.2$ cm, and a distance of closest approach along the $z$ axis of $< 0.4$ cm.
%A jet is called taggable if 
%it is matched to a jet formed from tracks, reconstructed using a simple cone-clustering algorithm of $\rm \Delta {\cal R} < 0.5$ 
%from tracks with $p_T>0.5$ GeV, each with at least one hit in the SMT detector,
%and each with a track distance of closest approach to the PV in the $x$-$y$ plane $<0.2$ cm, and a distance of closest approach 
%along the $z$ axis of $<0.4$ cm.  A minimum of two tracks is needed to form a track jet, and requires at least one track with 
%$p_{T} > $ 1 GeV.

%Then, the jets are tagged using a neural network $b$ tagging algorithm \cite{btag}.  The input variables to the 
%neural network include the impact parameters of all tracks in a track-jet,
%and information on secondary vertices in the jet. We apply a cut of $\rm > 0.775$ on this NN. 
All taggable jets are subjected to a neural network $b$ tagging algorithm \cite{btag} whose input variables 
include the DCA of each track in a jet
and information on secondary vertices in the jet. We define $b$-jet candidates by requiring that the neural network output be 
greater than $0.775$. 
%This algorithm selects about 50\% of $b$ jets with $p_T$ = 50 GeV, and only 1\% of light flavor ($u$, $d$, $s$ quarks or gluon) jets.
This algorithm selects about 60\% of $b$ jets with $p_T$ = 50 GeV, and only 1\% of light flavor ($u$, $d$, $s$ quarks or gluon) jets.

% corrections in the next section

%The taggability scale factors
%(Sec.\ref{sec:tagga_corr}) are applied to the MC jets to correct
%for the differences between data and MC taggability.
%\par
%In order to select the jets originating from heavy (b/c) quarks we use a Neural
%Network (NN) tagger. D\O\ NN tagger is a combination of JLIP, CSIP and
%SVT tagging algorithm and it enhances the efficiency for tagging the
%heavy quark jets~\cite{btagging-note}. The D\O\ BID group has measured the
%tagging rates for heavy/light quark jets in both data and MC for different operating
%points which correspond to different NN output values. In this analysis,
%we tag the jets using the ``TIGHT'' operating point which corresponds to
%NN output value of $\rm >0.775$. At the chosen NN operating point the b-jets
%selection efficiency is roughly 50\% for a jet $\rm p_T$ of 50 GeV. Since we
%directly tag the MC jets (based on their NN output values), we apply
%data-to-MC tagging correction factors to the MC jets (see Sec.~\ref{sec:btag_corr}).

\subsection{Missing Transverse Energy}
Neutrinos or other weakly-interacting neutral particles do not leave energy deposits 
in the detector.  Their presence is inferred from the 
measurement of significant \met\ in the event.
The missing transverse energy is determined from energies deposited in all calorimeter cells. 
%(including the
%Coarse Hadronic/unclustered energy cells) to estimate the
%(including unclustered energy).  We
%We correct this energy for the following:  coarse hadronic (CH) energy, JES, muon
%$p_T$, and electron and tau corrections.  The CH correction is necessary
%because the CH is noisy, and therefore not in the initial \met\ calculation.
%However, the CH energy in jets is real and must be added;  this is the CH
%correction.  The JES correction to \met\ subtracts the vector sum of the JES
The \met\ is corrected for JES, measured muon
$p_T$, electron and $\tau_h$ energy scales.  The JES corrected \met\ vector is 
obtained by adding the difference between the vector sums of uncorrected and JES corrected 
jet momenta to the uncorrected \met\ vector.  The muon correction 
reflects the fact that
muons deposit little energy in the calorimeter, and adjusts the \met\ for the $p_T$ of the 
muon.  Finally, electron and $\tau_h$ energy corrections are applied 
to the appropriate calorimeter cells in the \met\ calculation. 
%with appropriate adjustments 
%for jets, muons, electrons, and taus.  

%momenta carried away by the neutrinos.
% Then for all offline
%corrections e.g., JES correction, object smearing etc., the
%\met\  is corrected. 
%This analysis uses different cuts on
%the \met\ depending on whether the SM procethe final state.

%\subsection{Event Selection}

%We then select events requiring at least one lepton in all final states. This greatly reduces the number of final states considered, but also reduces the amount of difficult to model 
%QCD background processes. We also split the inclusive data set in to seven exclusive non-overlapping states \ref{sec:incstates}

\section{\label{sec:evgenpre}Modeling SM Predictions}

\subsection{\label{sec:evgen}SM Event Generation}
%\section{Standard Model event generation}
We generally estimate SM processes with MC generated events. A model-independent search incorporates many  
different processes to properly model the data. 
%We primarily use two generators for this purpose, {\sc alpgen} \cite{alpgen} for producing processes where we need to accurately incorporate jets produced in the hard scatter, and {\sc pythia} \cite{pythia} 
%where these are less important and our focus is on accurate hadronization and showering. When using {\sc alpgen}, we match the jets produced in the hard-scatter to {\sc pythia} for appropriate hadronization and showering. 
We use two generators for this purpose, {\sc alpgen} \cite{alpgen} for generation of all processes, except for diboson production which is generated with {\sc pythia} \cite{pythia}. {\sc pythia} is also used for hadronization 
and showering. 
%When using {\sc alpgen}, we match the jets produced in the hard-scatter to {\sc pythia} for appropriate hadronization and showering. 

{\sc alpgen} uses exact matrix elements at leading orders for QCD and electroweak interactions. The benefit of using {\sc alpgen} 
comes from its ability to calculate exact leading order terms for processes that include high jet multiplicities. 
{\sc alpgen} produces parton-level events with information on color and flavor, and can be matched to {\sc pythia} for parton evolution 
and hadronization.

Matching of a parton from {\sc alpgen} to {\sc pythia} showering has the fundamental difficulty of separation of the 
hard interaction from 
initial-state radiation (ISR) and final-state radiation (FSR). 
%We use the MLM matching scheme \cite{mlm_matching}, which looks for an appropriate $\Delta {\cal R}$ between 
%the partons from the matrix elements and the evolved jets, 
%rejecting those events without a match and also those with an additional unmatched jet, except in the sample of highest jet 
%multiplicity. 
To address this problem we use the MLM matching scheme \cite{mlm_matching}.  In
this scheme each final state parton from the
matrix element is matched in $\Delta {\cal R}$ to an evolved jet.  We
further reject events which contain an additional jet not
matched to a final state parton, except in the sample with the
highest number of final state partons.

%This eliminates double-counting from collinear partons with overlapping matrix elements and partons that are too soft to produce their own jet.
%All processes are generated with {\sc alpgen}, except for diboson production which is generated with {\sc pythia}.
%The diboson processes are generated with {\sc pythia}, while all other processes are generated using {\sc alpgen}.

The following processes are considered, where $j$ is a light jet ($g$,$u$,$d$, or $s$), $\ell$ is a lepton, $N$ is an integer $\geq 0$ and 
$lp$ represents a light parton:

\begin{enumerate}
\item$W + Nj$
\item$Z/\gamma^{*} + Nj$
\item$W + c\overline{c} + Nj$
\item$W + b\overline{b} + Nj$
\item$Z/\gamma^{*} + c\overline{c} + Nj$
\item$Z/\gamma^{*} + b\overline{b} + Nj$
\item$t\overline{t} \rightarrow (2\ell + 2\nu + 2b) + Nj$
\item$t\overline{t} \rightarrow (\ell\nu + 2b + 2lp) + Nj$
\item$WW$
\item$WZ$
\item$ZZ$
\end{enumerate}

%The factorization scale used for W + jets is the following:

%\begin{equation}
%Q^{2} = M_{W}^{2} + \sum_{jets}{p_{T}^{2}(j)}
%\end{equation}

%and for the Drell-Yan processes:

%\begin{equation}
%Q^{2} = M_{Z}^{2} + {p_{T}^{2}(Z)}.
%\end{equation}

The processes involving heavy flavor (HF) quarks ($c$ and $b$) are treated separately from light quark 
processes because they are often associated with particularly interesting final 
states, and we generate large number of MC events for these final 
states. Some of these processes are included in the light parton simulations, so we remove the events 
with heavy flavor quarks from the light-parton samples so as to avoid double-counting. 

For some objects, other programs provide more accurate simulations of their properties and decays. Specifically, {\sc tauola} 
\cite{tauola} is used for $\tau$ decays, and {\sc evtgen} \cite{evtgen} is used for the decay of $b$ hadrons.

We assume a mass of $172.5$ GeV for the top quark, consistent with recent measurements \cite{topmass}.

\subsection{\label{sec:detsim}Detector Simulation}
%\section{ Simulation}
The events produced from the above combination of generators are processed through the \dzero\ detector simulation and combined with random beam crossing  
events taken from data (Sec.~\ref{lumireweighing}). The detector simulation is based on {\sc geant} 3.2.1~\cite{geant}, 
to which two types of correction factors are applied.  The first type of correction is event reweighting, where 
an overall correction is applied to the MC event, rather than to the measured kinematic properties of reconstructed 
objects.  For example, we 
apply weights to account for the difference in reconstruction efficiencies between data and MC. 
%These weights depend on the kinematics of the particle.
%Object dependent weights are also
%sometimes applied to adjust MC when there are known problems with
%a given MC event generator. 
%In particular, we apply object scale
%factors to correct for electron and muon efficiencies, and we
%reweight the {\sc alpgen} $Z$ boson $p_T$ spectrum.  
%The other type of correction we apply is to modify the objects in a MC event to
%account for the fact that our detector resolution and energy scale is
%not as good as our simulation.  
Another type of correction modifies the objects in a MC event to account for the fact that the simulation has better 
resolution and a different energy scale than the detector.
These corrections generally depend on properties of the objects in an event.
The specific corrections used in this analysis are described below.
%For example, we apply jet
%shifting, smearing, and removal (JSSR) in the MC, and we smear the
%MC muon track $p_T$.

\subsubsection{Instantaneous Luminosity Reweighting \label{lumireweighing}}
%The MC uses random beam crossing events from data to model additional $p \bar{p}$ interactions in the event and detector noise. 
%#Zero-bias events are collected using  the trigger that fires on random beam crossings with only a prescale used to prevent saturation of the data acquisition system. 
%These events are added to the MC as background to the hard processes 
%modeled in the simulation. 
%#In early Run II, there were an average of 2.3 collisions in each beam crossing, and by the end of Run II, the average increased to 5.8. 
%As the random beam crossing events are collected for a different instantaneous luminosity profile 
%than the data, the MC is reweighted to match the instantaneous luminosity distribution 
%in data.

The trigger selecting random beam crossings records
data with a different instantaneous luminosity profile from that
of the triggers utilized to record the data used in this
search. A weight is introduced in the MC events to match the
instantaneous luminosity distribution in data.

\subsubsection{$Z_{\text PV}$ Reweighting}
Our simulated events have a narrower $z_{\text PV}$ distribution 
than is observed in data.  We therefore apply a weight to each 
event, based on the $z_{\text PV}$ of the event, to increase the 
relative weight of 
events farther from the center of our detector to match the observed distribution.

\subsubsection{JES \label{jes_corr}}
We apply JES corrections to jets in both data and MC \cite{jetcrosssection}.  The
purpose of the JES corrections is to correct the measured jet energy to 
that of the particles in the jet.  
Jet energies initially determined from the calorimeter cell energies do 
not exactly correspond to the energies of final state particles that traverse 
the calorimeter.  As a result, a detailed calibration is applied 
separately in data and MC.  In general, the energy of all final state 
particles inside the jet cone, $E_j^{\rm \/ptcl}$,
can be related to the energy measured inside the jet cone, $E_j$, by
$E_j^{\rm ptcl}$ $=(E_j-O)/( R\,S )$.
Here, $O$ denotes an offset energy, primarily from additional 
interactions in or out of time with an event.
$R$ is the average response of the calorimeter to the particles in a
jet, and $S$ is the correction factor for the net energy loss from particles that scatter out of 
or into the jet cone.  For a given cone radius, $O$ and $S$ are
functions of the jet $\eta$ within the detector.  $O$ is also a
function of the number of reconstructed event vertices and the
instantaneous luminosity;  $R$ is the largest correction factor and 
reflects the lower response
of the calorimeter to charged hadrons relative to electrons and photons.  It
also includes the effect of particle energy loss in front of the calorimeter.  The 
primary response correction is derived from studies of 
$\gamma+$jet events, and depends on jet energy and pseudorapidity.  For 
all jets that contain non-isolated muons, we add the muon momenta to that of the jet.  Under
the assumption that these muons are from semileptonic decays of $b$ 
quarks, we also add an estimated average neutrino momentum assumed to 
be collinear with the jet direction.
%The correction procedure discussed above does not correct all
%the way back to the original quark parton energy.

\subsubsection{Jet Shifting, Smearing, and Removal (JSSR)\label{jssr_corr}}
Additional corrections beyond the JES are needed to take into account threshold and resolution effects for jets. The JSSR 
corrections are determined from $Z/\gamma \rightarrow ee$ + 1
jet events.  The $Z/\gamma$ and the jet should be produced approximately 
back-to-back in $\phi$ with the same $p_T$.  This is quantified by a $p_T$ 
imbalance variable,
$\Delta S = \left(p_T^{j} - p_T^{Z/\gamma}\right)/p_T^{Z/\gamma}$.  For jets with a $p_T$ well above the 
reconstruction 
threshold, the distribution of $\Delta S$ is Gaussian in both data and 
MC.  The difference in the means of these distributions yields a shift 
that is applied to the MC jet energies to match the data, and a 
smearing is applied to MC jets based on the difference in the standard 
deviations of these distributions.  
Jets that fail the $p_T>20$ GeV requirement after shifting and smearing 
corrections are removed from further consideration.  
%For more details, see~\cite{d05609}.  To apply this correction, we use
%the RunJSSR class from caf\_mc\_util p18-br-77.

\subsubsection{Efficiencies \label{eff_corr}}
The efficiency of the MC simulation of our detector tends to be larger than the true efficiency of the 
detector.  To account for this, we introduce scale factors to adjust the MC efficiency to
match that observed in data.  The efficiencies for electrons and muons
are obtained using $Z \rightarrow e e$ and $Z \rightarrow \mu \mu$ events.  One
of the decay products of the $Z$ boson is the tag object, which is required to pass restrictive 
reconstruction requirements and be matched to an object
that could have fired the trigger for the event.  Object
efficiencies are then obtained using the second object from the 
$Z$ decay.  

%Details on these
%measurements can be found~\cite{d05105} and~\cite{d05157}.  

%For electrons,
%we use the preselection efficiency parametrization in the electron pseudorapidity measured by the calorimeter ($\eta_{det}$),
%and the efficiency parameterization in the 2-dimensional plane of the pseudorapidities and azimuthal angles ($\phi$) measured in the calorimeter.  For muons, we use the
%$\eta_{det}-\phi$
%parameterization of the muon outer detector efficiencies, with the
%tracking efficiency parametrization in the 2-dimensional plane of $z$ and pseudorapidity measured by the tracking system. For muons we also use the $p_T-N_{jet}$ dependent
%corrections to the isolation. 

%definition.  For all of these corrections,
%we use the spc files found in emid\_eff v7-preliminary-07, muid\_eff
%v04-01-03, and top\_cafe v01-05-02.

\subsubsection{Track $P_T$ Resolution\label{res_corr}}
Electron energies are measured in the calorimeter. However, energy deposition does not depend on the 
charge of the electron, which is determined by the curvature of the associated track in the magnetic field.  An 
incorrectly reconstructed track can therefore lead to an incorrect charge 
assignment. Bremsstrahlung from electrons can affect the curvature of the tracks.  Also, a soft interaction in 
the inner detector can result in the process $e^{+} \to e^{+}e^{-}e^{+}$, leading to charge misidentification if 
the wrong sign electron track is associated with the electron.
This difficulty is also present in tau decays when at least one hadron is produced.
%  The requirement that muon tracks in the central tracker be 
%matched to the muon system segments makes sign misidentification significantly lower for muons.

Because the rate of charge misidentification is not properly modeled in the detector simulation, we add a scale factor to 
electron and tau MC events to approximate the appropriate rate of charge mis-identification.  We determine this scale factor by using 
dielectron events consistent with $Z \rightarrow ee$ decays;  and we only consider events with dielectron invariant mass between 70 to 110 GeV to avoid biases against physics beyond the SM.

%The incorporation of this scale factor is difficult because a direct fit to the full data sample would bias us in our search for new physics. We therefore 
%restrict our sample to dielectron events that have invariant masses in the $Z$ boson peak. For this study, we are only looking at the electron calorimeter 
%energies, and we do not use the track \pt~ or look at the sign of the electrons. In the mass range used in this study, $60$ GeV $< M_{inv} < 120$ GeV, we see 
%very little contribution from multijet processes. We can therefore assume that the events in this region come exclusively from Drell-Yan production

The disagreement in track resolution between the data and MC also affects muon $p_T$ measurement,
%We smear our muon track $p_T$ using the ApplyMuonSmear class from  caf\_mc\_util p18-br-77.
which is corrected using smearing parameters determined by comparing the data and MC mass peaks for 
$Z \rightarrow \mu \mu$ and $J/\psi \rightarrow \mu \mu$ decays.
%Details can
%be found in~\cite{D05444}.

\subsubsection{Electron Energy Smearing}
%A different issue with data/MC agreement in electron final states is the electron $p_T$ resolution.  The calorimeter in the MC has slightly better 
%We also correct the electron $p_{T}$ resolution.  The calorimeter in MC has slightly better 
%The calorimeter in MC has better electron $p_T$ resolution than observed in the \dzero\ detector.  We correct this using 
In the simulation, the electron $p_T$ reconstructed in the calorimeter
has a better resolution than in the data.  We correct this using 
a Gaussian smearing function tuned to reproduce the shape of the $Z \to ee$ peak.

\subsubsection{Jet Taggability \label{sec:tagga_corr}}
The jet taggability rates (Sec.~\ref{btaggingdetails}) are found to be different for MC and data.
To correct for this difference, correction factors are applied as scale factors depending on $p_T$, $\eta$ and $z_{\text PV}$ of the jet \cite{higgs}. 
%are adapted from the search for
%$WH\rightarrow \ell \nu b\bar{b}$ \cite{higgs}, and 
%~\cite{whanalysis_note}.
%The said corrections have been derived inclusively (independent of
%jet flavor) in two in sets of events (a) $ |z_{\rm{PV}}|<30$ cm and
%(b) $\rm 30<|z_{PV}|<60$ cm. Therefore, 
%are applied as 

\subsubsection{$b$-tagging Rate\label{sec:btag_corr}}
As detailed in Sec.~\ref{btaggingdetails}, we apply
%As detailed above, we apply
a tagging algorithm to both data and MC jets to select jets originating from
heavy ($b$/$c$) quarks. However, the algorithm can select mistagged light jets.  
The tagging rates (for both heavy and light parton
jets) depend on the $p_T$ and $\eta$ of the jets. The heavy-quark tagging rates are measured separately
in both data and MC using dedicated samples.
%~\cite{btagging-note}.
The performance of the $b$-tagging algorithm
in MC events is better than in data.
To correct the tagging rates in MC events, we first
determine the flavor of the tagged jet by matching it in $\rm\Delta {\cal R}$
with the initial parton. Depending on the flavor of the jet, we apply a
per-jet scale factor given by ${\rm SF} ={\epsilon^{data}}(p_T,\eta)/{\epsilon^{MC}} (p_T,\eta)$, where 
${\rm \epsilon^{data}}(p_T,\eta)$ and ${\epsilon^{MC}}(p_T,\eta)$ are the $b$-tagging efficiencies
for a given parton flavor for data (MC) events.  To maintain correct normalization, a small downward 
correction is applied to non-$b$-tagged jets.

%It is to be noted here that there is no such
%data-to-MC scale factors applied to the light jets which are mistagged.

\subsubsection{Weak Gauge Boson $p_T$\label{zpt_corr}}
%The MC method of using {\sc alpgen} matched to {\sc pythia} is inconsistent with data 
%for the $p_T$ spectra of the $Z$ and $W$ boson at small values of $p_T$. The $Z$ boson $p_T$ is therefore reweighted to match the 
%measured $Z$ boson $p_T$ distribution from $Z \rightarrow ee$ decays \cite{zpt}.
The $p_T$ distribution of the $Z$ boson from {\sc alpgen} MC is corrected to match the distribution 
observed in data in $Z \rightarrow ee$ decays \cite{zpt}.
A modified reweighting is carried over to the $W$ boson $p_T$ based on the theoretical ratio of the $W$ to $Z$ $p_T$ spectra 
\cite{boson_ptspectra}.

\subsubsection{$\Delta \phi$ \label{delphi_corr}}
%There is disagreement between data and MC in distributions of $\Delta \phi$  between objects. The main reason is the fact that 
%leading order MC
%that we are using does not describe this distribution, and additional next to leading order corrections are needed.
%Large 
%discrepancies are seen between leptons 
%in dilepton final states, and between the lepton and $\met$ in single-lepton + jets final states. We assume that 
%these distributions are not due to contributions from new physics, but rather indicate a modeling deficiency. 
We apply a $\Delta \phi$-dependent weight specifically for this analysis using the inclusive distributions 
described in Sec.~\ref{sec:incstates} to correct the $\Delta \phi$ between leptons in dilepton final states and 
the lepton and $\met$ in single-lepton + jets final states.
%With that assumption, we  apply a reweighting scheme to the $\Delta \phi$ distributions.  This reweighing 
%is calculated specifically for this analysis using the inclusive distributions described in Section~\ref{sec:incstates}.  
This reweighting 
affects not only the $\Delta \phi$ distributions, but also other quantities that depend on the angular distribution of particles  
such as the $p_T$ of the $W$ boson.

%\subsubsection{\met\ correction\label{met_corr}}
%Corrections to \met\ are done using the ReComputeMET class of caf\_util.  
%Before
%corrections, the \met\ is based solely on the raw calorimeter cell energy with no
%minimum $E_T$ or $\eta$ cuts.  We
%correct this energy for the following:  coarse hadronic (CH) energy, JES, muon
%$p_T$, and electron and tau corrections.  The CH correction is necessary
%because the CH is noisy, and therefore not in the initial \met\ calculation.
%However, the CH energy in jets is real and must be added;  this is the CH
%correction.  The JES correction to \met\ subtracts the vector sum of the JES
%corrections for the jets from \met\ .  The muon correction reflects the fact that
%muons deposit little energy in the calorimeter by subtracting the associated
%muon track $p_T$ from the \met\ .  Finally, the electron and tau corrections
%are there because both objects have energy corrections applied to them, which
%must be taken into account by the \met\ .
%~\cite{Note4474}.

%D0Geant and corrections
%for generator shortcomings ( event weights) and detector simulation 
%shortcomings (object weights).  
%Must describe how weights are obtained and show plots illustrating
% what the corrections do.

%\subsection{\label{sec:misid}Use of Data for Misidentified Objects}
%\input misident.tex

\section{\label{sec:incstates}Inclusive Final States}
The seven inclusive non-overlapping final states are specified in Table \ref{tab:mis_final_states} by the relevant objects 
and their selection criteria. The 
additional objects ($X$ in the table) are selected as shown in Table \ref{tab:mis_additional_objects}.  Events with photons are 
rejected, mainly due to difficulties in modeling.
The seven states ($e$ + jets, $\mu$ + jets, $ee$, $\mu \mu$, $\mu e$, $e \tau$, $\mu \tau$ ) were each selected to correspond 
to a specific SM process. 
%We treat the Drell-Yan (D-Y) contributions to the $ee$ and $\mu\mu$ final states without light partons separately from those with 
%light partons because it improves agreement between data and MC.  
\ctable[
  caption = {Inclusive final states and their object selections, where $p_T^{\text {min} }$ is the 
minimum allowed value of $p_T$ and $| \eta |^{\text {max} }$ is the maximum allowed value of $|\eta|$.},
  mincapwidth = 2.5in,
  pos = {htp},
  label = {tab:mis_final_states}
]
{rccc}
{
%\tnote[a]{ $p_T^{\text {min} }$ minimum allowed value for the $p_T$}
%\tnote[b]{ $| \eta |^{\text {max} }$ maximum allowed value for the $|\eta|$}
\tnote[a]{$X$ $\neq$ $e$, $\mu$, $\tau$, $\gamma$}
\tnote[b]{$X$ $\neq$ $e$, $\mu$, $\tau$, $\gamma$}
\tnote[c]{$X$ $\neq$ $\mu$, $\tau$, $\gamma$}
\tnote[d]{$X$ $\neq$ $e$, $\tau$, $\gamma$}
\tnote[e]{$X$ $\neq$ $\tau$, $\gamma$}
\tnote[f]{$X$ $\neq$ $\gamma$}
\tnote[g]{$X$ $\neq$ $e$, $\gamma$}
}
{
\hline
\hline
%Final State & Object & $p_T^{\text {min} }$ (GeV)\tmark[a] & $| \eta |^{\text {max} }\tmark[b]$ \\
Final State & Object & $p_T^{\text {min} }$ (GeV) & $| \eta |^{\text {max} }$ \\
\hline
\multirow{3}{*}[11.5pt]{$e$ + jets + $X$\tmark[a]} & $e$ & 35 & 1.1\\
 & jet & 20 & 2.5\\
 & \met & 20 & -\\
%\hline
\multirow{3}{*}[11.5pt]{$\mu$ + jets + $X$\tmark[b]} & $\mu$ & 25 & 1.5\\
 & jet & 20 & 2.5\\
 & \met & 20 & -\\
%\hline
\multirow{1}{*}{$ee$ + $X$\tmark[c]} & $e$ & 20 & 1.1\\
%\hline
\multirow{1}{*}{$\mu\mu$ + $X$\tmark[d]} & $\mu$ & 15 & 1.5\\
%\hline
\multirow{2}{*}[5.75pt]{$\mu e$ + $X$\tmark[e]} & $\mu$ & 15 & 1.5\\
 & $e$ & 15 & 1.1\\
%\hline
\multirow{2}{*}[5.75pt]{$e \tau$ + $X$\tmark[f]} & $e$ & 15 & 1.1\\
 & $\tau$ & 15 & 1.1\\
%\hline
\multirow{2}{*}[5.75pt]{$\mu\tau$ + $X$\tmark[g]} & $\mu$ & 15 & 1.5\\
 & $\tau$ & 15 & 1.1\\
%\hline
\hline
\hline
}

\ctable[
  caption = {Criteria required for inclusion as additional objects ($X$) in one of the seven final states listed in Table \ref{tab:mis_final_states}.},
  mincapwidth = 2.5in,
  pos = {htp},
  label = {tab:mis_additional_objects}
]
{ccc}
{
}
{
\hline
\hline
Object & $p_T^{\text {min} }$ (GeV) & $| \eta |^{\text {max} }$ \\
\hline
$e$ & 15 & 1.1\\
%\hline
$\mu$ & 15 & 1.5\\
%\hline
$\tau$ & 15 & 1.1\\
%\hline
jet & 20 & 2.5\\
\hline
\hline
}

\begin{itemize}

\item{$e$ + jets}

The electron + jets final states have more background from multijet events, where a jet is misidentified as an electron, than the other electron final 
states.  Therefore the likelihood criterion used is tighter than in other final states, $\mathcal{L}_{e} > 0.95$.  
We also require at least one jet having $E_{T} > 20$ GeV, $\met > 20$ GeV, and an $e$ $p_T > 35$ GeV.  This 
final state is dominated by $W$ + jets events with $W \to e \nu$ decays.
%is tighter than the default $top~tight$ definition.

\item{$\mu$ + jets}

The $\mu$ + jets final state is dominated by $W$ + jets events with $W \to \mu \nu$ decays. 
%This state is defined by exactly one muon with $p_{T} > 25$ GeV and $|\eta| < 1.5$. 
To reduce the amount of multijet 
background, at least one jet having $E_{T} > 20$ GeV is required, as well as \met $> 20$ GeV and a muon with $p_{T} > 25$ GeV. 
Just as the $e$ + jets final state, this final state is inclusive in jets with no other additional objects allowed.

\item{$ee$}

The dielectron final state requires each electron to have $p_{T} > 20$ GeV and $\mathcal{L}_{e} > 0.85$. The electrons are also restricted to be in the 
central calorimeter, $|\eta|<1.1$, and the jets have the same criteria as for the other final states.  This final state is dominated by $Z/\gamma^{*} \to ee$ events.
%The end calorimeters were excluded for this analysis because of inconsistencies between  
%electrons measured in the central calorimeter and electrons measured in the end caps when attempting to fit histograms in the dielectron final state 
%normalization fit.

\item{$\mu \mu$}

The dimuon final state requires at least two muons with the muon-$p_{T}$ criteria lowered to $p_{T} > 15$ GeV because of the smaller contribution from 
multijet background. 
Any jet must have $p_{T} > 20$ GeV. This final state is inclusive in both jets and muons, but an additional $e$ or $\tau$ lepton 
places the event in the $\mu e$ or $\mu\tau$ final states.  Analogous to the $ee$ channel, this final state is dominated by $Z/\gamma^{*} \to \mu \mu$ events.

\item{$\mu e$}

The $\mu e$ final state is inclusive except for $\tau$ leptons; $e \mu \tau$ events are assigned to the $e \tau$ final state.   
This final state is dominated by $Z/\gamma^{*} \to \tau \tau$ events.
%The electrons can be identified normally as stated in section~\ref{sec:object_id} or as misidentified $\tau$'s with the electron 
%separation $NN_e < 0.2$.  This final state is dominated by $Z/\gamma* \to \tau \tau$ decays.

\item{$e \tau$}

The $e\tau$ sample is inclusive in all objects. The electron and $\tau_h$ $p_{T}$ are required to be at least 15 GeV. 
The electron likelihood is set to $L_e > 0.95$ to reduce the large multijet background as many apparent $\tau_h$ correspond to  
misidentified jets. 
%The hadronic $NN_h$ cuts are the same as $\mu \tau$, but 
The parameter that separates electron from hadronic taus, $\rm NN_e$, is set to 0.8 to reduce the contribution from dielectron events.  This final state is 
also dominated by $Z/\gamma^{*} \to \tau \tau$ events.
%Also, $\tau$-type 2 candidates are required to be outside of an EM calorimeter $\phi$ crack.

\item{$\mu \tau$}

The $\mu\tau$ state contains at least one muon and one $\tau_h$. It is inclusive in all objects except 
electrons, whose presence would move the event to the $e \tau$ final state.  This final state is also 
dominated by $Z/\gamma^{*} \to \tau \tau$ events.
%The requirements are $\mu$ $p_{T}>15$ GeV and $\tau$ $p_{T} > 15$ GeV. 
%The $\tau$ $NN_{h} > 0.9$, and the $\tau$ of type 2 has an additional electron separation cut of $NN_{e} > 0.2$.

\end{itemize}

\section{Inclusive Normalization Fits}

Our model does not provide proper normalization of different MC contributions because, for example, of higher-order corrections needed for the 
leading-order or leading-logarithm cross section calculations.  To avoid 
uncertainties in normalization, we perform a fit, described below, for each of the inclusive final states to obtain scale factors that reproduce the distributions of 
the selected data using a combination of the SM MC and multijet predictions determined from data. We treat the Drell-Yan (D-Y) contributions to the $ee$ and $\mu\mu$ final states without light partons separately from those with 
light partons because it improves agreement between data and MC.  
%As there are separate reconstruction 
%efficiencies for $b$-tagged jets, we fit the $W$ and $Z$ events with heavy flavor (HF) jets separately
%Since the contributions from heavy flavor is expected to be small to the inclusive final states, so we fit for the heavy flavor to light parton (hf/lp) content instead of fitting of fitting for heavy flavor alone.  
 % Since the seven states are non-overlapping, they can be combined as an input to the {\sc vista} algorithm without fear of double-counting.

The fits for normalization factors are performed on kinematic distributions of different object quantities, 
altering the overall normalization of each input process contributing to final state so that the $\chi^{2}$ probability for that final state is 
minimized for the combined fit.  To avoid fitting to data at the highest values of $p_T$, where new physical processes can be 
important, we only use events that are not in the high $p_T$ tail, which contains 10\% of the events.
%check each object in the event to see if the object $p_T$ is outside the bulk of the distribution, where we have 
%set the $p_T$ cuts for the tail such that 90\% of events will fall into the bulk.  
Distributions of basic quantities such as 
$\met$, $p_{T}$, $\eta$, $\Delta \phi({\text {obj}},~ \met$) of leptons 
and jets (here obj refers to the momentum of the object considered) are used in the fits while more complex 
variables are used to check the quality of the overall 
fit. 
The latter variables include the mass or transverse mass 
$M_T = \sqrt{(p_{T,1}+p_{T,2})^{2} -(\vec{p}_{T,1} +\vec{p}_{T,2})^{2} }$ 
of two or more objects, jet multiplicities, and the $p_T$ of the $W$ and 
$Z$ bosons. 
If an event contains any object outside the $p_T$ range defined above, then none of the objects in the event 
are used in the fit.

% A full list of the processes which are normalized based on these inclusive fits, and the final states and number of 
% data events that are used to determine their values, are shown in Table \ref{tab:mis_fit_processes}.
The list of the seven final states, the processes that are normalized through the inclusive fits to each of the final states, and
 the number of events in each final state are shown in Table \ref{tab:results_scale_factors}.
 Once the fitted values are extracted, the distributions are rescaled accordingly, and the total 
background contribution, $B$, for a particular final state is 
\begin{equation}
 B = \sum_i^{N_{bkg}} S_{i} B_{i}
\end{equation}
 where the scale factor ($S_{i}$) for each background process ($B_{i}$) is determined from the final state in which its 
 contribution is most important and that scale factor is used in all other final states to which that 
background contributes.  $N_{bkg}$ refers the total number of all the SM processes contributing to a 
particular final state.
 
A simplified example for the $e$ + jets + $X$ final state ($X \neq e$, $\mu$, $\tau$, $\gamma$) 
is used to illustrate the procedure.  
The $e$ + jets + $X$ state is dominated by $W \to e \nu$ events, but there is a significant contribution from multijet and 
Drell-Yan events.  We use the normalization factor for the Drell-Yan process, determined through a separate fit to the $ee$ + $X$ final 
state ($X$ $\neq$ $\mu$, $\tau$, $\gamma$), in the $e$ + jets fit.  We also fix the scale factors to one for rare processes which have 
contributions that are too small to fit accurately in $e$ + jets, such as the $t\bar{t}$ contribution. 
We then fit for the SM $W$ boson and multijet contributions in the data.  The fit optimizes agreement between the distributions in data and 
the SM prediction for the variables listed above.  The result of the fit is two overall weights, one for $W\to e \nu$ and 
one for ${\rm multijet} \to e + {\rm jets}$.
% We fit the given histograms by varying
% the $W$ and multijet contributions; the result are two scale factors,
% the overall weights for the $W\to e \nu$ and ${\rm multijet} \to e + {\rm jets}$ 
% contributions.
 %Then, the $W$ and multijet contributions will find the best agreement to 
 %fit the given histograms and two scale factors will be used to give an overall 
 %weight to the $W \rightarrow e \nu$ and ${\rm multijet} \rightarrow e +{\rm jets}$ contributions. 

%The fit itself minimizes the negative logarithm of the likelihood function for each set of parameters.
% and converts this value to a $\chisq$. 
The distributions of the variables for the input processes are not varied, only their relative contributions. 
The fit is performed using the {\sc minuit} program \cite{minuit}. 
%It minimizes the $\chisq$ of the fitting histograms by looking at the differences, bin-by-bin, between 
%the data and the SM background. The floating parameters are modified until a minimum is found. 
%Only two or three parameters for each of the final states are modified. 
For single-lepton states and hadronic $\tau$ final states, multijet events are a significant background. We assume that the 
contribution from other SM processes modeled by the MC samples to the multijet background is small. 
The scale factors of input processes for the MC events should also account for the contributions of the processes to the multijet background. 
%The main effect of contributions from physics processes modeled by the MC to the multijet background would be that the multijet state would resemble the process MC, making it difficult for the fit to reliably find the 
%multijet contribution.
The main effects of contributions from any of the MC processes to the multijet background would be to decrease the scale factor for 
backgrounds modeled by MC.
%, and would reduce our sensitivity to new physics in {\sc vista} due to contributions from new physics to the 
%multijet background.

\begin{figure}[htp]
\begin{center}
%\subfigure[] {
\includegraphics[width=3.0in]{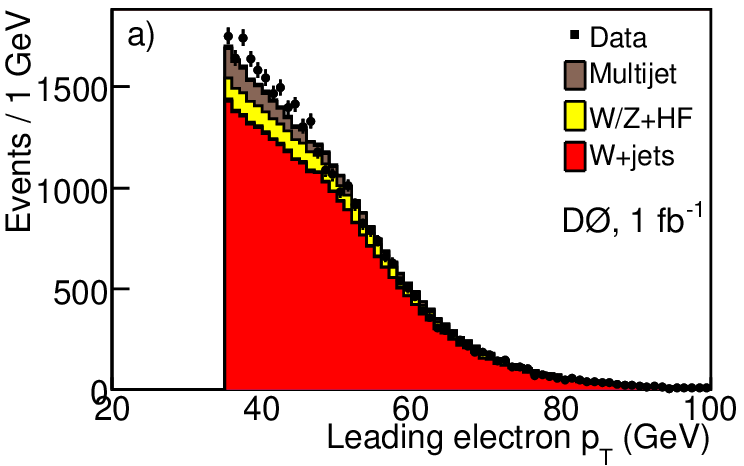}
%}
%\subfigure[] {
\includegraphics[width=3.0in]{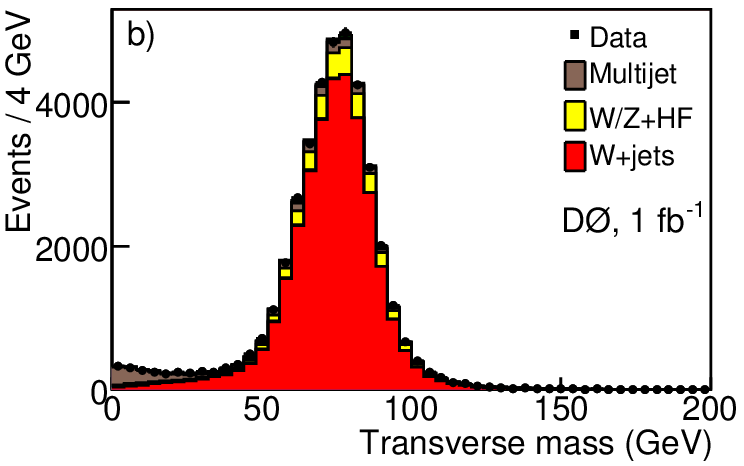}
%}
\end{center}
\caption{(color online) $e$ + jets final state (a) electron $p_T$ histogram  and (b) transverse mass ($e,\met$) check histogram. }
\label{fig:mis_ejetsstates_b}
\end{figure}

\begin{figure}[htp]
\begin{center}
%\subfigure[] {
\includegraphics[width=3.0in]{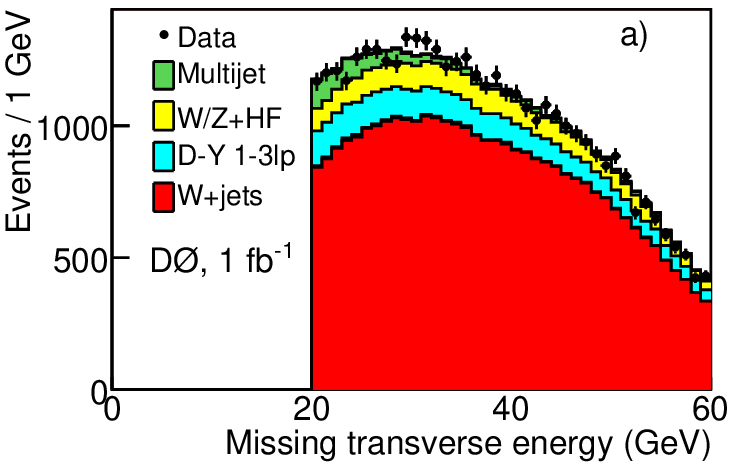}
%}
%\subfigure[] {
\includegraphics[width=3.0in]{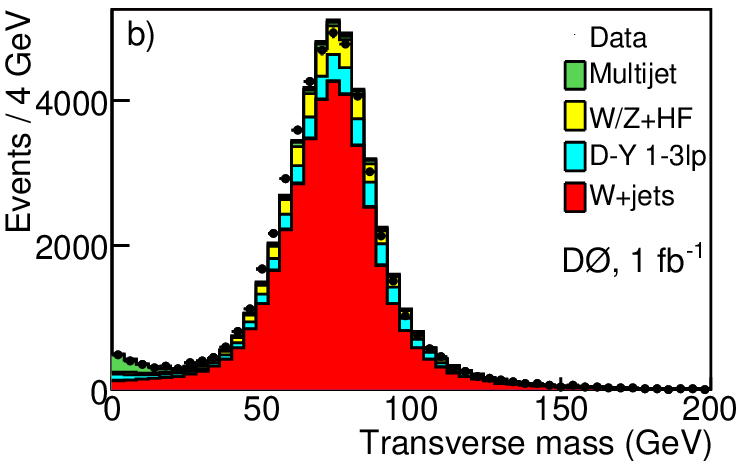}
%}
\end{center}
\caption{(color online) $\mu$ + jets final state (a) $\met$ histogram  and (b) transverse mass ($\mu$, $\met$) check histogram.}
\label{fig:mis_mujetsstates_b}
\end{figure}

\begin{figure}[htp]
\begin{center}
%\subfigure[] {
\includegraphics[width=3.0in]{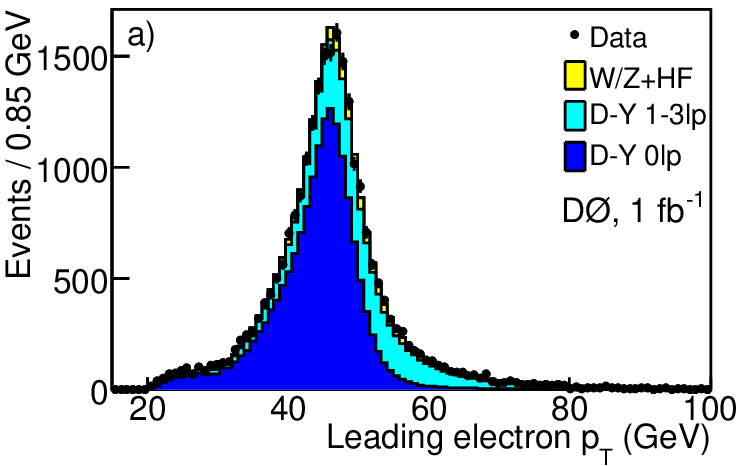}
%}
%\subfigure[] {
\includegraphics[width=3.0in]{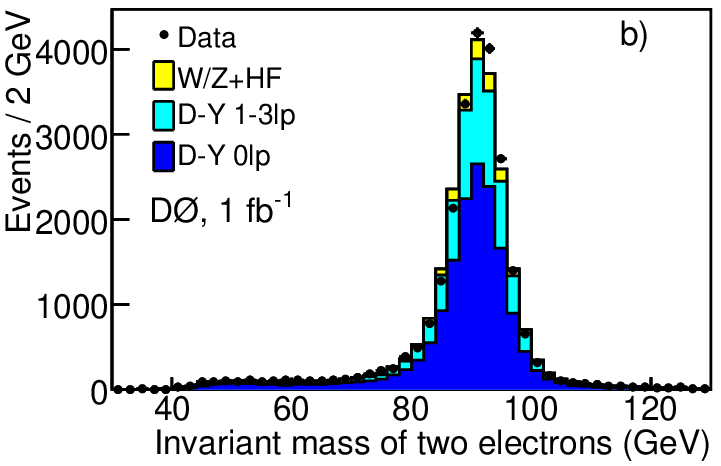}
%}
\end{center}
\caption{(color online) $ee$ final state (a)  leading electron (with highest $p_T$) $p_T$ fit histogram and (b) invariant mass ($e$,$e$) check histogram.}
 \label{fig:mis_eestates_b}
\end{figure}

\begin{figure}[htp]
\begin{center}
%\subfigure[] {
\includegraphics[width=3.0in]{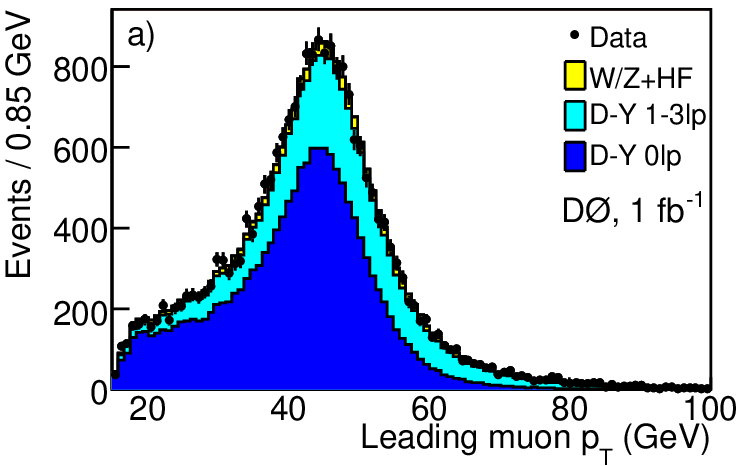}
%}
%\subfigure[] {
\includegraphics[width=3.0in]{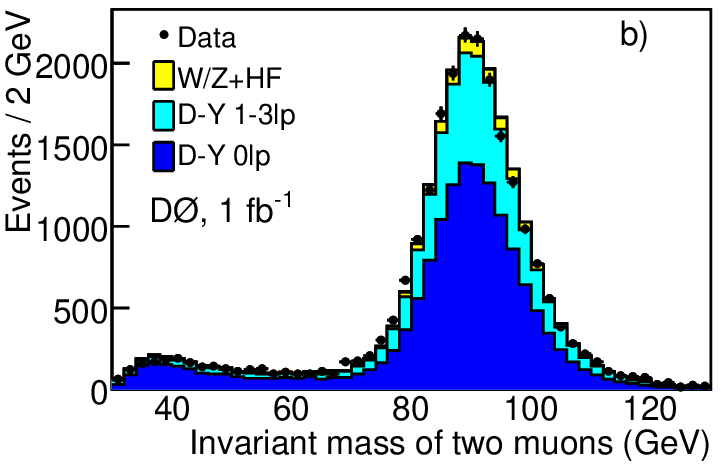}
%}
\end{center}
\caption{(color online) $\mu \mu$ final state (a) leading muon $p_{T}$ fit histogram and (b) invariant mass ($\mu,\mu$) check histogram.}
\label{fig:mis_mumustates_b}
\end{figure}

\begin{figure}[htp]
\begin{center}
%\subfigure[] {
\includegraphics[width=3.0in]{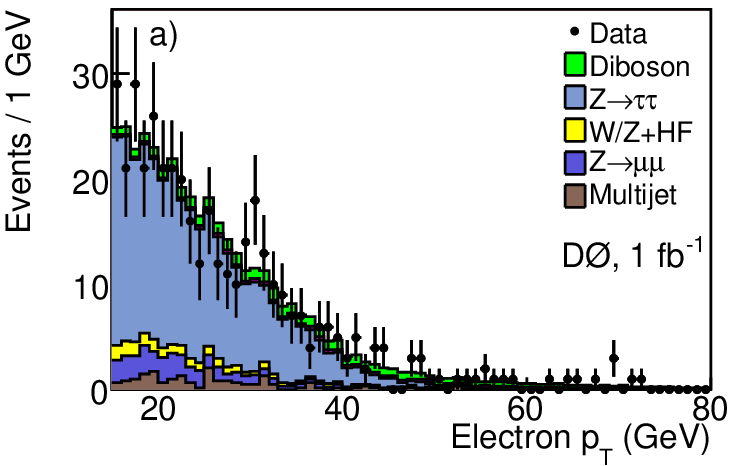}
%}
%\subfigure[] {
%\includegraphics[width=3.0in]{used_figures/mis/mue_ept0.eps}
\includegraphics[width=3.0in]{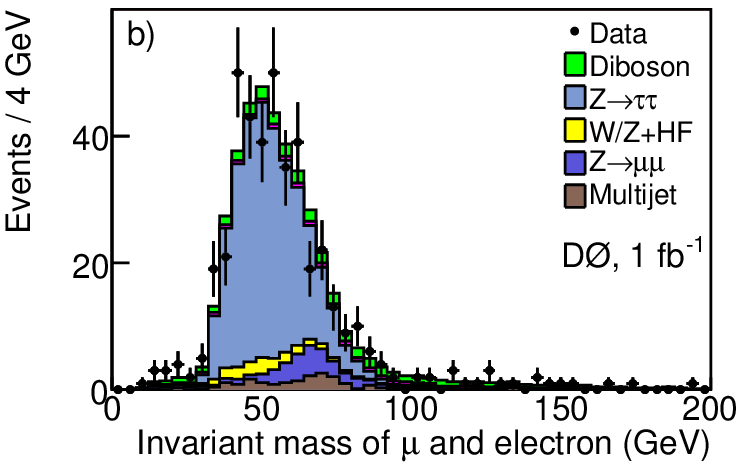}
%}
\end{center}
\caption{(color online) $\mu e$ final state (a) electron $p_T$ fit histogram and (b)  invariant mass ($\mu,e$) check histogram.}
\label{fig:mis_muestates_b}
\end{figure}

\begin{figure}[htp]
\begin{center}
%\subfigure[] {
%\includegraphics[width=3.0in]{used_figures/mis/etau_taueta.eps}
%}
%\subfigure[] {
\includegraphics[width=3.0in]{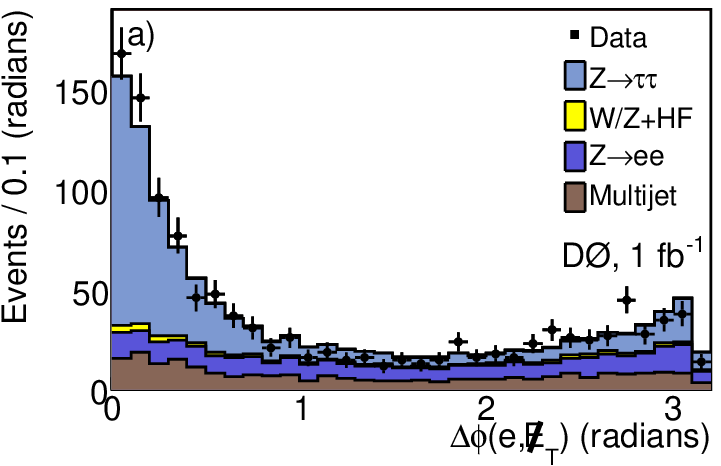}
\includegraphics[width=3.0in]{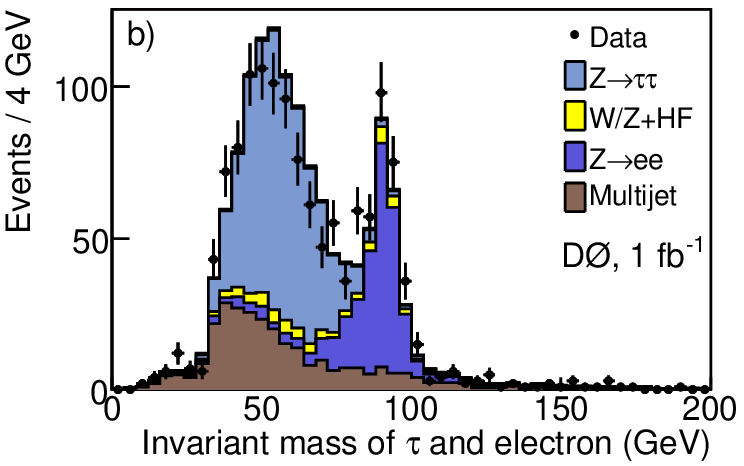}
%}
\end{center}
%\caption{(color online) $e\tau$ final state (a) tau $\eta$ fit histogram and (b) invariant mass ($e$,$\tau$) check histogram.}
\caption{(color online) $e\tau$ final state (a) The $\Delta\phi$($e$,$\met$) fit histogram and (b) invariant mass ($e$,$\tau$) check histogram.}
\label{fig:mis_etaustates_b}
\end{figure}

\begin{figure}[htp]
\begin{center}
%\subfigure[] {
\includegraphics[width=3.0in]{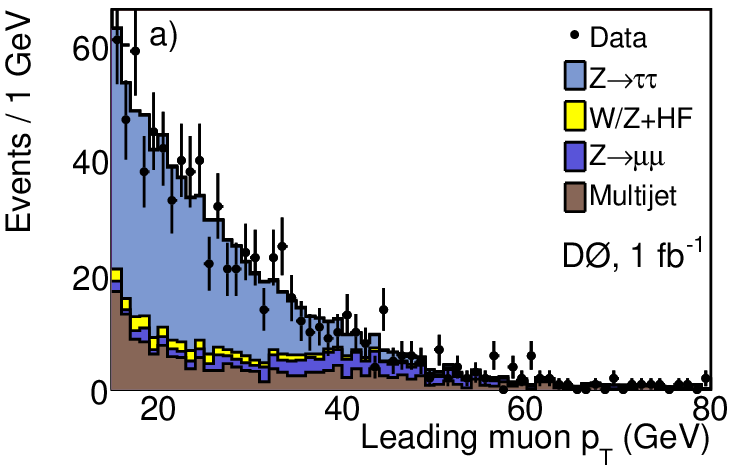}
%}
%\subfigure[] {
\includegraphics[width=3.0in]{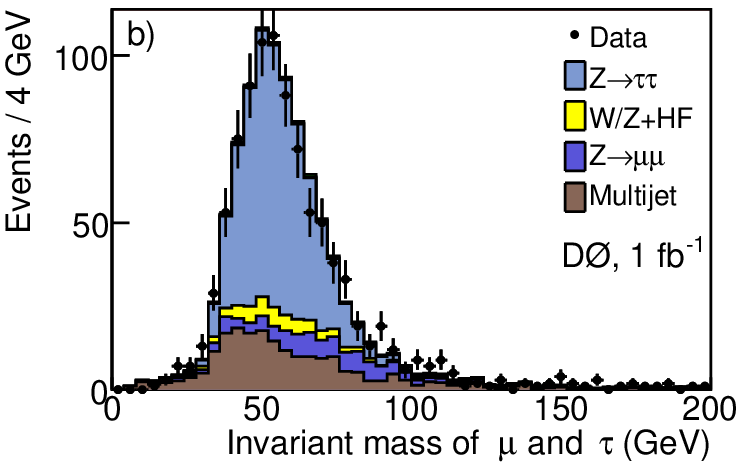}
%\includegraphics[width=3.0in]{used_figures/mis/mutau_taupt0.eps}
%}
\end{center}
\caption{(color online) $\mu \tau$ final state (a) muon $p_T$ fit histogram  and (b) invariant mass ($\mu, \tau$) check histogram.}
\label{fig:mis_mutaustates_b}
\end{figure}

The main purpose of the normalization process is to assure that the fundamental SM processes are well-modeled.
The results of the fit are then checked for qualitative agreement with the data. 
%A Kolmogorov-Smirnov (KS) 
%probability is 
%determined for each of the histograms to provide a quantitative check for comparison. Additionally, 
The overall scale factors are checked to compare to those from dedicated analyses. If the normalization factors 
are properly included in the MC, then all the scale factors should equal unity.  
%No specific cut is required for the KS probability because the main quantitative analysis will be done at the later {\sc vista} and 
%{\sc sleuth} stages. 
One histogram that is included in the overall fit and one check histogram that is not part of the fit are 
shown for each of the seven 
final states in Figs. \ref{fig:mis_ejetsstates_b}  --  \ref{fig:mis_mutaustates_b}. In the figures, 
the leading and second electron are the electrons with highest $p_T$ in the event and next highest $p_T$ in 
the event, with a similar definition for leading and second muons and jets. 
%\ref{fig:mis_mujetsstates_b}  --  \ref{fig:mis_muestates_b}.  

The electron $p_T$ distribution in Fig.~\ref{fig:mis_ejetsstates_b} shows a clear disagreement between data 
and simulation 
in this kinematic region arising from the need for a large multijet contribution at low $p_T$, and other variables that provide better agreement with a 
smaller multijet contribution.  
However, the discrepancy at low $p_T$ should not mask the presence of new physics at high $p_T$, which is the main focus of this 
analysis.
%\ref{fig:mis_ejetsstates_check}, \ref{fig:mis_mumustates_fit}, \ref{fig:mis_mumustates_check}, 
%\ref{fig:mis_eestates_fit},  \ref{fig:mis_eestates_check}, \ref{fig:mis_mutaustates_fit}, \ref{fig:mis_mutaustates_check}, \ref{fig:mis_etaustates_fit}, \ref{fig:mis_etaustates_check}, 
%\ref{fig:mis_muestates_fit}, \ref{fig:mis_muestates_check}.

%All of the processes considered in each of the seven final states with the number of events are shown in Table \ref{tab:results_scale_factors}.
%, and the $\chisq$ of the fit are shown in Table \ref{tab:results_scale_factors}.

\ctable[
  mincapwidth = 2.5in,
  caption = {The contributions used in the inclusive fits for each of the inclusive final states and the number of data events in 
each final state.  The dominant SM process is listed first for each final state.  In the $e \tau$ and $\mu 
\tau$ final states, the multijet background also includes a contribution from $W$ + jets.},
  pos = {htp},
  label = {tab:results_scale_factors}
]
{p{2.5cm}p{2.5cm}c}
{
}
{
\hline
\hline
State & SM process & Events \\
\hline
%\multirow{3}{*}[11.5pt]{$e$ + jets + $X$} & $W \rightarrow e \nu$ & \multirow{3}{*}[11.5pt]{40k} \\
% & multijet $e$ & \\
% & hf/lp ratio & \\
\multirow{3}{*}[11.5pt]{$e$ + jets + $X$} & $W$ + jets & \multirow{3}{*}[11.5pt]{40k} \\
 & Multijet & \\
 & $W/Z$ + HF & \\
%\multirow{3}{*}[11.5pt]{$\mu$ + jets + $X$} & $W \rightarrow \mu \nu$ & \multirow{3}{*}[11.5pt]{50k}\\
% & multijet $\mu$ & \\
% & hf/lp ratio & \\
\multirow{3}{*}[11.5pt]{$\mu$ + jets + $X$} & $W$ + jets & \multirow{3}{*}[11.5pt]{50k}\\
 & Multijet & \\
 & $W/Z$ + HF & \\
%\multirow{3}{*}[11.5pt]{$ee$ + $X$} & $Z \rightarrow  e e$ + 0lp & \multirow{3}{*}[11.5pt]{25k}\\
% & 1-3lp/0lp ratio &  \\
% & hf/lp ratio &  \\
\multirow{3}{*}[11.5pt]{$ee$ + $X$} & D-Y + 0lp & \multirow{3}{*}[11.5pt]{25k}\\
 & D-Y 1-3lp ratio &  \\
 & $W/Z$ + HF &  \\
%\multirow{3}{*}[11.5pt]{$\mu \mu$ + $X$} & $Z \rightarrow  \mu \mu$ + 0lp & \multirow{3}{*}[11.5pt]{24k}\\
% & 1-3lp/0lp ratio & \\
% & hf/lp ratio & \\
\multirow{3}{*}[11.5pt]{$\mu \mu$ + $X$} & D-Y + 0 lp & \multirow{3}{*}[11.5pt]{24k}\\
 & D-Y + 1-3 lp & \\
 & $W/Z$ + HF & \\
%\multirow{3}{*}[11.5pt]{$\mu e$ + $X$} & $Z \rightarrow  \tau \tau$ & \multirow{3}{*}[11.5pt]{0.34k}\\
% & $W$+jet / multijet $\tau$  & \\
% & hf/lp ratio & \\
\multirow{3}{*}[11.5pt]{$\mu e$ + $X$} & $Z \rightarrow  \tau \tau$ & \multirow{3}{*}[11.5pt]{0.34k}\\
 & Multijet & \\
 & $W/Z$ + HF & \\
%\multirow{3}{*}[11.5pt]{$e \tau$ + $X$} & $Z \rightarrow  \tau \tau$ &\multirow{3}{*}[11.5pt]{1.3k}\\
% & $W$+jet / multijet & \\
% & hf/lp ratio & \\
\multirow{3}{*}[11.5pt]{$e \tau$ + $X$} & $Z \rightarrow  \tau \tau$ &\multirow{3}{*}[11.5pt]{1.3k}\\
 & Multijet & \\
 & $W/Z$ + HF & \\
%\multirow{3}{*}[11.5pt]{$\mu \tau$ + $X$} & $Z \rightarrow  \tau \tau$ & \multirow{3}{*}[11.5pt]{1.0k}\\
% & $W$+jet / multijet & \\
% & hf/lp ratio & \\
\multirow{3}{*}[11.5pt]{$\mu \tau$ + $X$} & $Z \rightarrow  \tau \tau$ & \multirow{3}{*}[11.5pt]{1.0k}\\
 & Multijet & \\
 & $W/Z$ + HF & \\
\hline
\hline
}

%\section{Text File Production}

%Once all of the normalization weights are determined and the input processes checked for agreement, the input files for the {\sc vista} and {\sc sleuth} algorithms are created. 
%These algorithms take text file inputs which only contain the most basic information about the objects. The overall event weight, run/event number, and vertex position are kept along with the object $p_{T}$, $\eta$, 
%and $\phi$. Using this simple information, the algorithms quantify the overall agreement between Data and MC.

%The text files are created in the same way that the histograms were created for the fit. The same computer code is used in their production, with the addition of one input weight that comes from the inclusive 
%state normalization fits.
%The text files are created using the same algorithm that was used to create the histograms for the fit, with the addition of one input weight that comes from the inclusive state normalization fits.

%An example of one line of a $\mu \tau$ text file can be seen in Figure \ref{fig:mis_textevt}.

%\begin{figure}[htp]
%\begin{center}
%\includegraphics[width=3.0in]{used_figures/mis/text_event.eps}
%\end{center}
%\caption{(color online) The figure shows one line of a $\mu \tau$ text file used as input into the {\sc vista} algorithm. Only the run and event numbers, the vertex position, weight, and the object \pt , $\eta$, $\phi$ 
%information are kept for each event.  In the figure, each object is shown in a different color.}
%\label{fig:mis_textevt}
%\end{figure}

\section{\label{sec:exstates}Exclusive Final States}
After determining the normalization scale factors, the seven inclusive subsets are merged to create an input file for the {\sc vista} algorithm~\cite{cdfPRD}. 
Each MC and background event 
is given a weight calculated from the data based scale factors and any required corrections. The {\sc vista} algorithm, developed by the 
CDF Collaboration, is a tool that performs a broad check of the agreement between data and the SM.  We modified the CDF algorithm for 
our analysis strategy as described above. The resultant {\sc vista@D0 } algorithm focuses on the D0 high 
$p_T$ data to determine whether the data can be adequately described by the SM or if significant discrepancies can be confirmed.  
{\sc vista} mainly examines discrepancies that affect the overall distributions rather than narrow regions of phase space, 
addressing the numbers of expected events and MC/data agreement across full distributions of chosen variables. 
%Since {\sc vista} looks at many final states and histograms, the sensitivity to any individual discrepancy is reduced. This makes it only sensitive to relatively significant discrepancies. 

The use of standard object identification criteria (Sec.~\ref{sec:object_id}) provides 
great simplification in the analysis as data can be partitioned into exclusive
final states. The events are separated into homogeneous subsets of events according to the objects
contained in each event, resulting in 117 exclusive final states. Examples of such exclusive final states include 
 $\mu^{\pm} \tau^{\mp}$ + 2 jets + $\met\ $, 
$e^{\pm} \mu^{\mp}$ + 2 jets + $\met\ $, $e^{+} e^{+}$ + 3 jets, and $\mu$ + 4 jets + \met.

%Based on event weights from the correction factors, 
{\sc vista} performs two types of checks: first, it does
a normalization-only check on the number of events in each
exclusive state; the goodness of the fit is calculated using 
Poisson probabilities. Second, it calculates a
Kolmogorov-Smirnov statistic (and resulting fit probability) for
the consistency of all the kinematic distributions in any final state with the predicted SM distributions.
Both of these results require additional interpretation because
of the large number of trials (number of final states and/or the number of distributions) involved. 
When observing many final states, some disagreement is expected from statistical fluctuations in the data.  Thus the Poisson probability used 
to determine agreement is corrected to reflect this multiple testing.  A similar effect
occurs when comparing kinematic distributions, and again the probabilities
are first converted to standard deviations
and then corrected for the number of distributions examined.

Another algorithm we use to search for new physics is called {\sc sleuth}~\cite{run1sleuth2},
developed at the \dzero\ experiment during Run I (1992-1996) of the Tevatron.
{\sc sleuth} is an attempt to 
systematically search for new physics as an excess at the largest values  
of $\sum p_T$.  This variable corresponds to the sum of the values of the scalar $p_T$ of all objects in the event, including the $\met$. The 
{\sc sleuth} algorithm is quasi-model independent, where ``quasi'' 
refers to the assumption that the 
physics beyond SM will appear as an excess of events at large $p_T$.  Therefore {\sc sleuth} is expected to be most sensitive to
high-mass objects decaying into relatively few final-state particles.

For {\sc sleuth}, the {\sc vista} exclusive  0 and 1-jet final states are merged, as are the 2 and 3-jet final states, and 
%first {\sc vista} exclusive channels are combined by charge conjugation (so $e^+$X and $e^-$X are combined), and 
light-lepton universality is
assumed, combining $eX$ and  $\mu X$ channels.  Making these assumptions greatly reduces the number of states 
considered in {\sc sleuth} relative to {\sc vista}, and thus the trials factor, improving the statistical sensitivity by diminishing the chance of observing a large 
fluctuation.
Next, the $\sum p_T$ distribution in each channel is scanned to find a
cutoff that maximizes the significance of any excess in data relative to the SM
background, defining a lower bound for the $\sum p_T$ selection.  Finally, the probability 
for consistency with the SM of the largest values of the $\sum p_T$ 
is corrected for the number of possible lower bounds in any distribution, and
subsequently for the number of final states examined by {\sc sleuth}. This corrected probability corresponds to the
probability that any individual final state would yield 
probabilities as small as observed. 
%{\sc sleuth} is designed to search the highest $p_T$ values, and therefore is an ideal
%tool for evaluating an excess corresponding to physics beyond the SM.
%We have followed CDF by defining
We define
a significant output from {\sc sleuth} as one with a corrected
probability of $<~0.001$ (that is over 3 Gaussian standard deviations from the SM prediction using a 
one-sided confidence interval).  

\section{Sensitivity Test}

To check the sensitivity of a search with {\sc sleuth}, we examine whether a top quark (produced in $t \overline{t}$ pairs)
which contributes objects with high $p_T$ would have been discovered in the current data sample.
For this test, we used all the background samples,
except for the $t \overline{t}$ MC.  The main concern is whether other final states 
would compensate for the missing $t \overline{t}$ events, and thus {\sc sleuth} 
would not be sensitive to $t \overline{t}$ production in data.

\begin{figure}[htp]
\begin{center}
%\subfigure[] {
%\label{fig:vista_sleuth_ttbar_a}
%\includegraphics[height=3.0in,angle=90]{used_figures/vista_sleuth/sleuth_withtop.eps}
\includegraphics[width=3.0in]{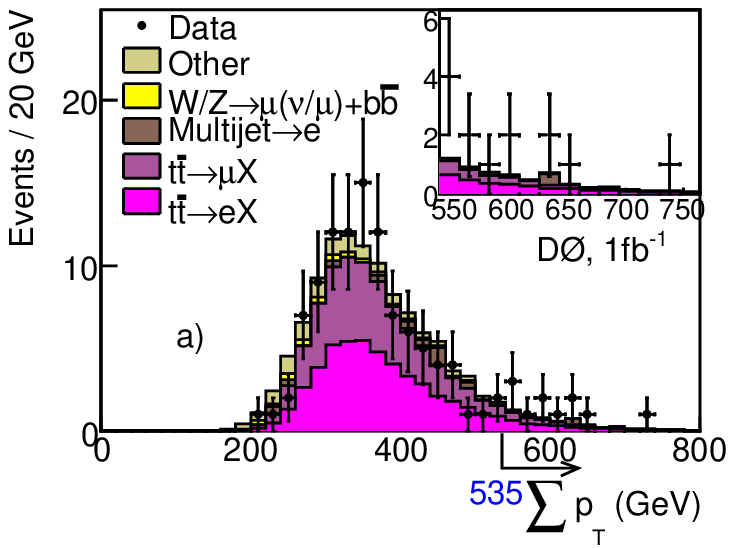}
%}
%\subfigure[] {
\includegraphics[width=3.0in]{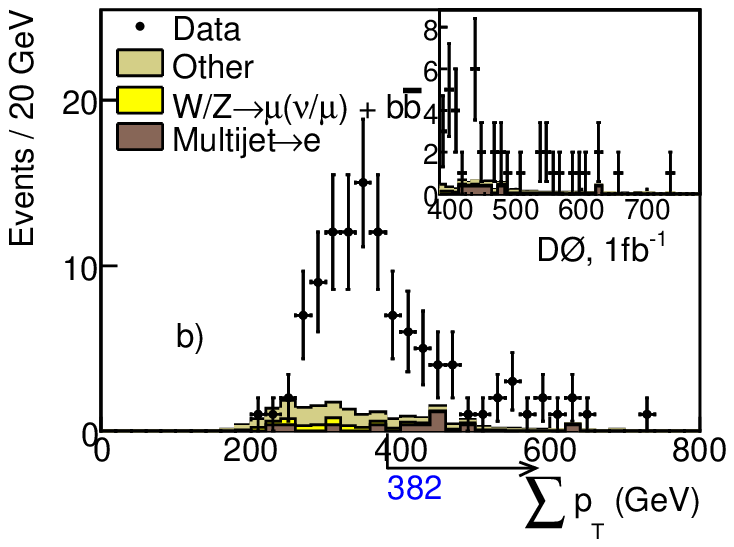}
%}
\end{center}
\caption{(color online) \label{fig:vista_sleuth_ttbar} Sensitivity to new physics test using the $t \overline{t}$ final state. (a) The 
$t\overline{t}$ MC is included, yielding only minor differences between data and SM background. The statistical agreement between the data and MC
for the distribution shown on inset is nearly 2 $\sigma$.
(b) The results of the entire analysis without the $t \overline{t}$ MC. In this case, {\sc sleuth} passes the
criterion of interest at 0.001 for this final state. The insets shows the distribution beyond the $\sum p_T$ 
cutoff.  ``Other" refers to contributions too small to list, including $W + b \overline{b} \to e \nu b 
\overline{b}$ events, $W+ c \overline{c} \to \ell \nu c \overline{c}$ events, $W +lp \to \ell \nu+ lp$ 
events, and diboson events.}
\end{figure}

We examine the $\ell jj b\bar{b} \met$ final state, which we expect to be dominated by $t \bar{t}$ events.  
Figure~\ref{fig:vista_sleuth_ttbar} shows that presence or absence of a $t \bar{t}$ signal has 
a great impact.  With a 
threshold of 0.001, the {\sc sleuth} test, including the $t \overline{t}$ MC,
yields a statistical probability of compatibility of 0.98 after correcting for the number of trials.  However, without the $t \overline{t}$ 
contribution this 
probability is $< 1.1 \times 10^{-5}$.   In Fig.~\ref{fig:vista_sleuth_ttbar} and other 
{\sc sleuth} plots, the insets show the results for data and MC that pass the $\sum p_{T}$ cut maximizing the significance of excess in data.

\section{\label{sec:results}Results}
\subsection{\label{sec:results_vista}Numerical discrepancy using the {\sc vista} analysis}

\begin{figure}[htp]
\begin{center}
\includegraphics[width=3.0in]{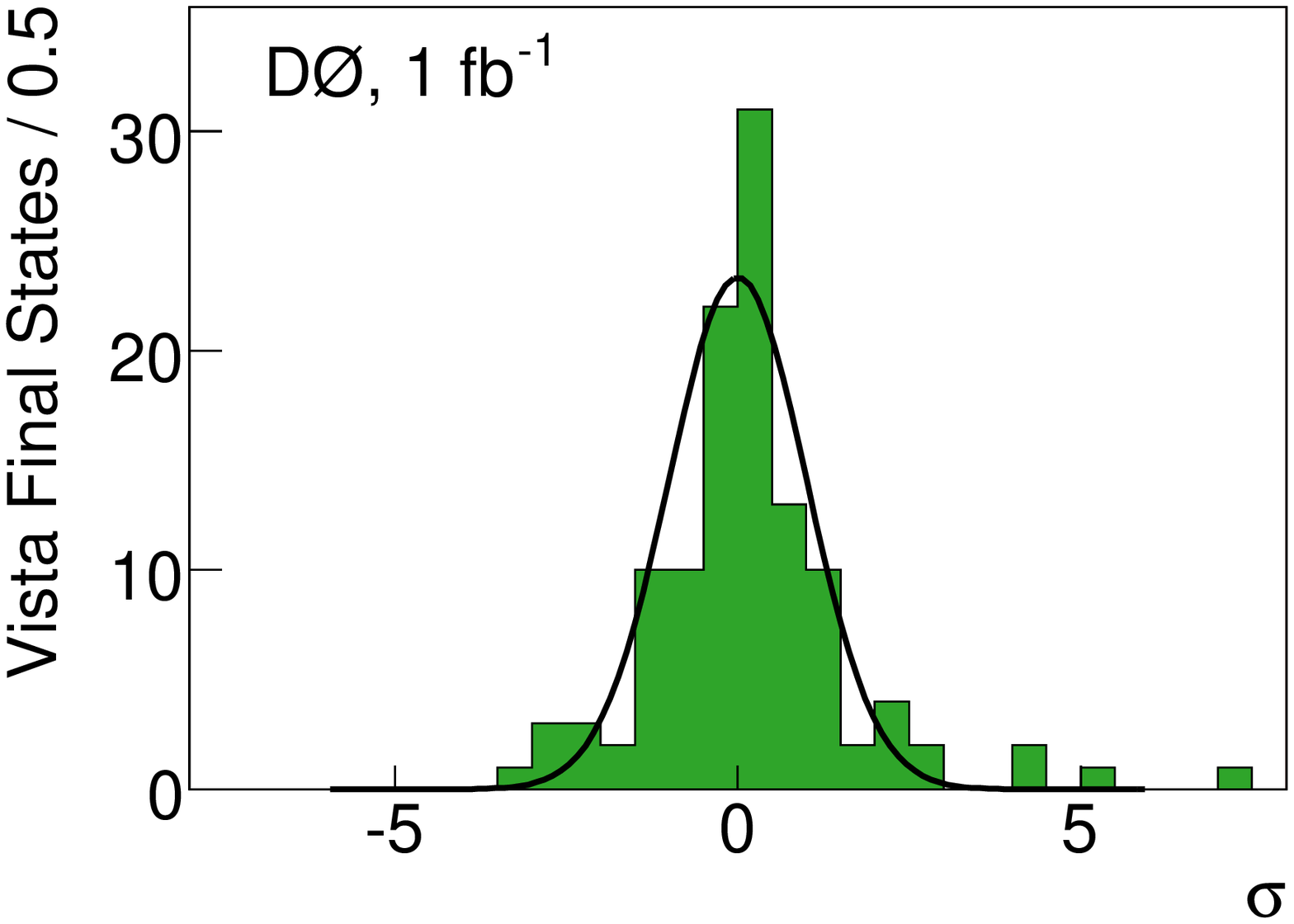}
\caption{(color online) \label{fig:results_vista_norm100pc} Distribution of discrepancies for the 117 final states relative to the SM in terms of standard deviations calculated in 
{\sc vista} final state before accounting for the trials factors. The curve represents a Gaussian distribution 
centered at zero to guide the eye. The distribution is expected to obey Poisson statistics, which is the reason the distribution is narrower than the Gaussian.}
\end{center}
\end{figure}

In {\sc vista}, the separation of the input data into final states completely defined by the objects in an event, yields a total of 117 unique exclusive final states. The 
probability ($\tilde{\mathcal{P}}$) that the yield observed in data results from a statistical fluctuation of the SM sample in 
channel $f_s$ is determined from 
\begin{equation}
\tilde{\mathcal{P}} = 1 - (1 - p_{fs})^{N_{fs}} \overset{p_{fs} \ll 1}{\approx} N_{fs} \times p_{fs}
\end{equation}    
where $N_{fs}$ is the number of trials and $p_{fs}$ is the probability that the number of events predicted for the channel $f_s$ in the SM would 
fluctuate to what is observed in data, 
before applying the correction for the number of trials.  The number of trials is $N_{fs} = 117$, corresponding to the number of final states, and 
\begin{equation}
p_{fs}=\int_0^{\infty} \exp\left[-\frac{(N-N_{B})^2}{2 \sigma^2_{B}}\right] dN \sum_{N_{\rm data}}^{\infty} \frac{N^i}{i!}e^{-N},
\end{equation}
where $N_{B}$ and $\sigma_{B}$ are the expected SM event 
yield from background and its uncertainty, respectively, and  $N_{\rm data}$ is the number of events observed in any channel.
The Gaussian significance is the value of $\sigma$ that satisfies the equation
\begin{equation}
\int_{\sigma}^{\infty}\frac{1}{\sqrt{2 \pi}}e^{- \frac{x^{2}}{2}}dx = \tilde{\mathcal{P}}. 
\end{equation}
The final state probabilities converted into standard deviations, before the correction factor for the number of trials, are shown 
in Fig. \ref{fig:results_vista_norm100pc}. This distribution shows most final states near $\sigma=0$, with some excess for $\sigma>3$.
%, where $\sigma$ being the standard deviation. 
Of the 117 final states, two show significant discrepancy after correction for the number of trials. These are the final states $\mu$ + 2 jets + 
$\met$, with a probability corresponding to a 4.5 $\sigma$ discrepancy, 
and $\mu^{+} \mu^{-}$ + $\met$ with a discrepancy of 6.7$\sigma$ (also shown in Fig~\ref{fig:results_vista_norm100pc}).

\begin{figure}[htp]
\begin{center}
%\subfigure[] {  
%\label{fig:results_countdiscrepant_a} 
%\includegraphics[height=3.0in,angle=270]{used_figures/results/vista_mu2jetsmet_jet2eta.eps}
%\includegraphics[height=3.0in,angle=270]{used_figures/results/plots_withCuts_2j1mu+1pmiss_jet2deteta.eps}
\includegraphics[width=3.0in]{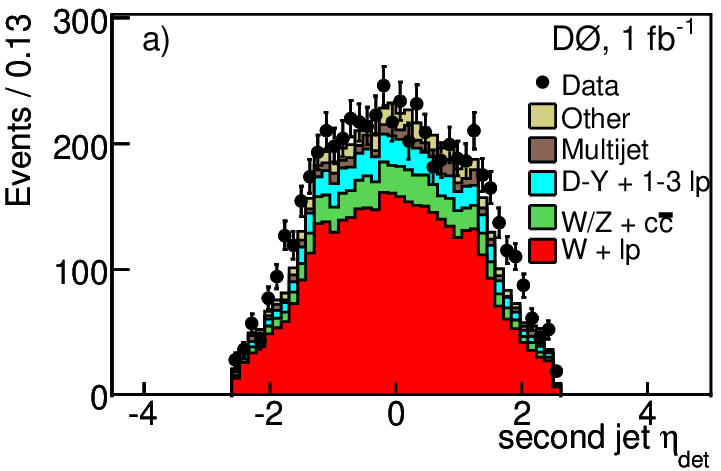}
%} 
%\subfigure[] { 
%\includegraphics[height=3.0in,angle=270]{used_figures/results/vista_mumumet_dphimumet.eps}
%\includegraphics[height=3.0in,angle=270]{used_figures/results/plots_withCuts_1mu+1mu-1pmiss_dphimumet.eps}
\includegraphics[width=3.0in]{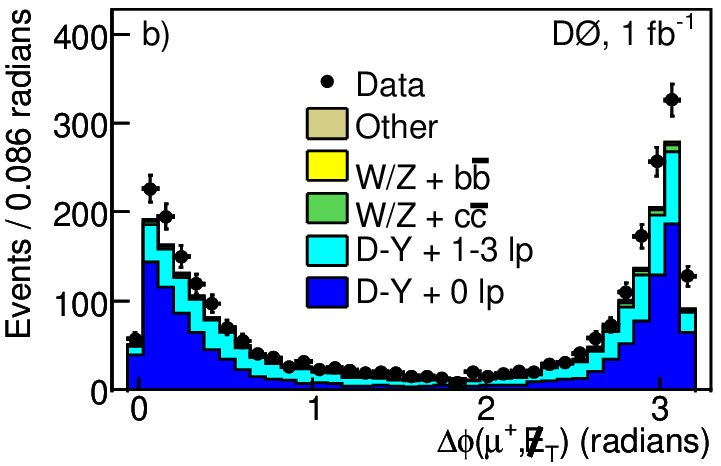}
%}
\end{center}
\caption{(color online) \label{fig:results_countdiscrepant} (a) The distribution of the pseudorapidity 
of the second jet with respect ot the center of the D0 detector in the $\mu$ + 2 jets + $\met$ channel. 
``Other" contains distributions too small to list individually, $ W + b \bar{b}$, diboson, $t \bar{t}$, and D-Y + 0 $lp$.
(b) The $\Delta \phi$ distribution between the $\mu^+$ and the $\met$ for the $\mu^{+}\mu^{-}+\met$ final state. 
``Other" contains distributions too small to list individually, diboson and $t \bar{t}$.} 
\end{figure}

%The $\mu$ + 2 jets + $\met$ final state discrepancy shows an excess of events with a muon at $\eta > 1.0$ as seen in Figure \ref{fig:results_countdiscrepant_a}.The excess points to an oversimplification in our 
%approach to trigger efficiencies. The proportion of events selected by single muon vs. muon plus jets triggers changes significantly as we increase jet multiplicity. These triggers introduce $\eta$-dependent 
%efficiencies which are not properly incorporated into our simple fits. 
The discrepancy for the $\mu$ + 2 jets + $\met$ final state shows the greatest difference from the SM prediction in the modeling of jet distributions.  There is a significant excess in the number of jets at 
high $|\eta|$, which points to likely problems with modeling ISR/FSR jets in the forward region, as can be 
seen in Fig. \ref{fig:results_countdiscrepant}a.  
This difference is observed in dedicated analyses \cite{dedicated}, and the discrepancy becomes less severe 
when using \sherpa\ \cite{sherpa} MC events.  

%A study of the track curvature of muons in data and in MC, and the associated resolution suggests 
%that an additional smearing should be applied in the MC to appropriately simulate 
%very high-$p_T$ muons.
The $\mu^{+} \mu^{-}$ + $\met$ discrepancy can be attributed to difficulties modeling the muon momentum distribution for high $p_T$ 
muons.  As noted in Sec.~\ref{res_corr}, the muon smearing modeling is based on muons from $Z$ and $J/\psi$ decays, dominated by muons 
below 60 GeV, and is not as reliable at high $p_T$.  
The prime signature of poorly simulated high $p_T$ muons is an excess of $\met$ because of the mismodeling of the resolution of the mismeasured track. The 
$\Delta \phi$ between the positive muon and $\met$ in the $\mu^{+} \mu^{-}$ + $\met$ final state is shown in Fig. 
\ref{fig:results_countdiscrepant}b, where the excess tends to be for events where the \met\ is collinear with a muon.
%As the 
%two muons tend to be produced back-to-back in these events, events 
%with the missing energy pointing toward one muon usually means muon momentum mismeasurement.

\begin{figure}[htp]
\begin{center}
\includegraphics[width=3.0in]{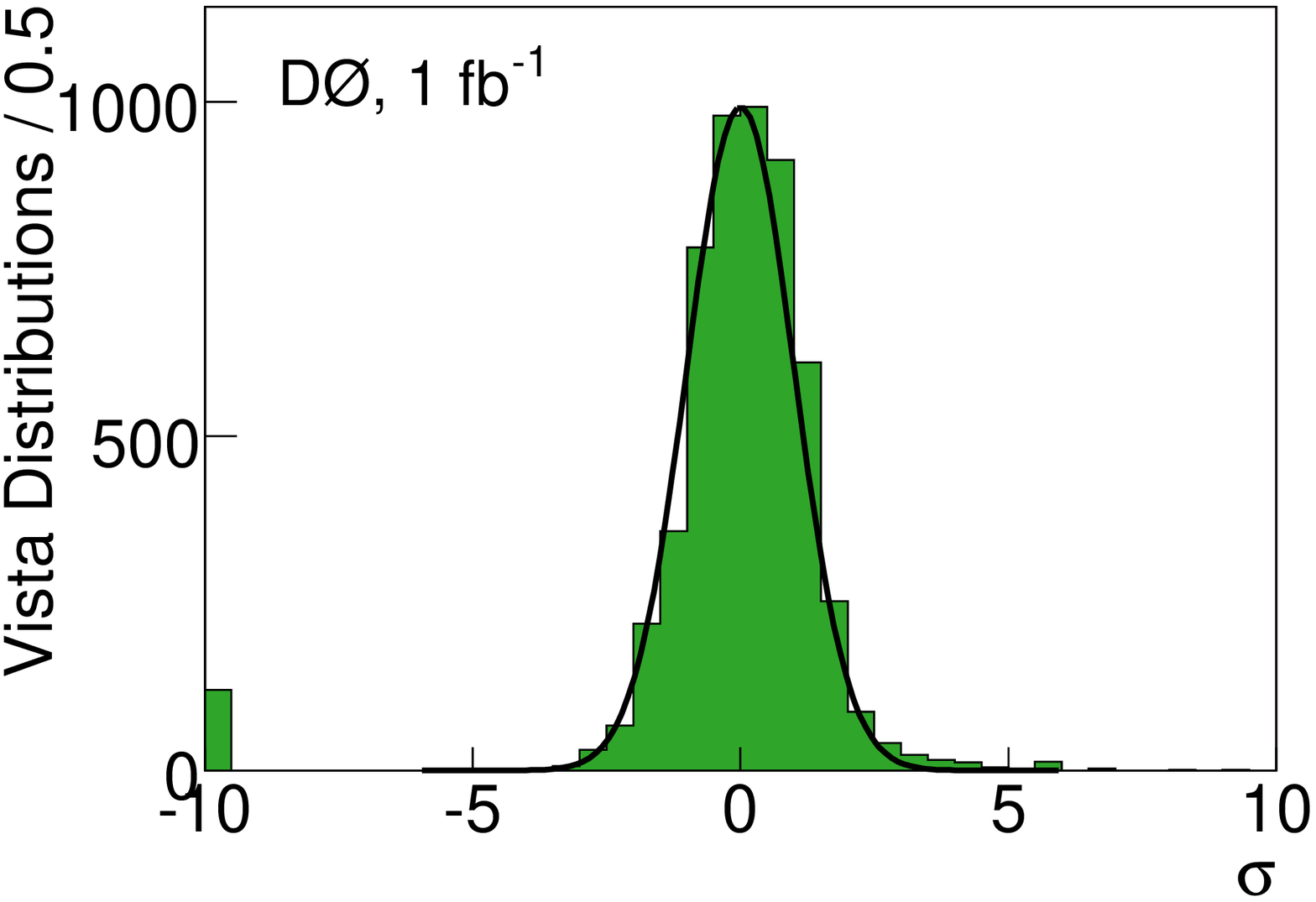}
\caption{(color online) \label{fig:results_vista_shape100pc} The $\sigma$ distribution for the 5543 {\sc vista} comparisons before accounting for number if the trials.  The curve represents a Gaussian distribution 
centered at zero to guide the eye.
There are 116 distributions in the underflow bin with $\sigma \leq -10$.  This is expected as histograms with KS probabilities $>0.99999$ are rounded to 1, and 
appear in the underflow bin.}
\end{center}
\end{figure}

\subsection{\label{sec:results_vista_shape}{\sc vista} Shape Analysis of Discrepancies in Distributions}
The 117 final states contribute a total of 5543 individual one-dimensional distributions in various variables, 
and comparison between simulation and data is performed for each. The trials-factor adjusted probability is determined from 
$\tilde{\mathcal{P}} = 1 - (1 - 
p_{\text{shp}})^{5543}$, where $p_{\text{shp}}$ is the Kolmogorov-Smirnov (KS) probability to observe a discrepancy for any individual distribution (before applying the correction for 5543 trials). 
As with the probability for a final state normalization discrepancy in any final state, the 
probability for a discrepancy in a spectrum is converted into units of standard deviation. Any deviation \textgreater 3$\sigma$ is considered discrepant. The 
distribution of deviations before correction for the number of trials is shown in Fig. \ref{fig:results_vista_shape100pc}. 
%This distribution approximates a slightly shifted Gaussian of the expected width, but several 
%distributions appear in the tails. 
%The shift to the right (toward poor agreement) is expected because we introduce scale factors only for the most important discrepancies (minor systematic discrepancies are not 
%individually treated, and these contribute preferentially towards bad agreement).

\begin{figure}[htp]
\begin{center}
%\subfigure[] {
\label{fig:results_shapediscrepant_mismu2j_a} 
\includegraphics[width=3.0in]{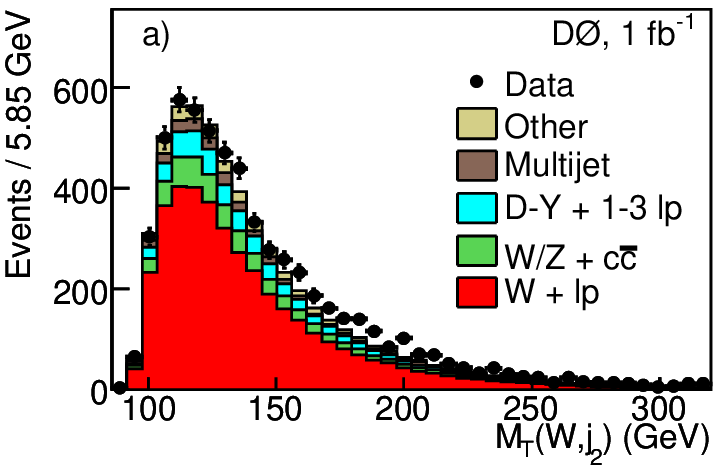}
%}

%\subfigure[] { 
\label{fig:results_shapediscrepant_mismu2j_b} 
\includegraphics[width=3.0in]{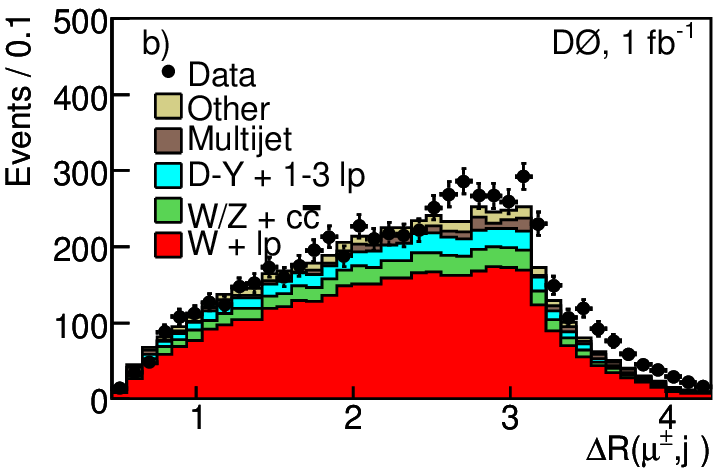}
\label{fig:results_shapediscrepant_mismu2j_c} 
\includegraphics[width=3.0in]{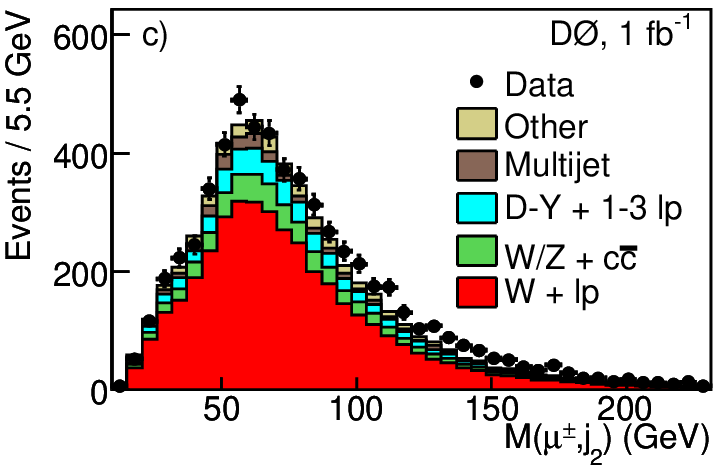}
\includegraphics[width=3.0in]{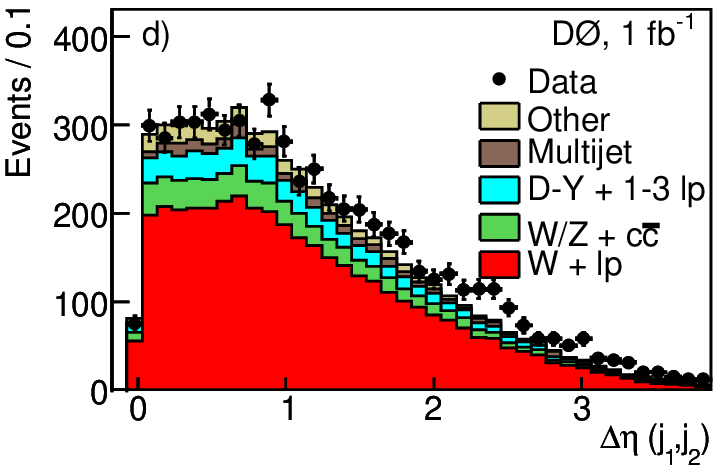}
\label{fig:results_shapediscrepant_mismu2j_d} 

%} 
\end{center}
%\caption{(color online) The plot \ref{fig:results_shapediscrepant_a} shows the $\Delta \eta$ distribution for the 2 jets and 
%\ref{fig:results_shapediscrepant_b} shows the $\rm \Delta {\cal R}$ difference between the $\mu$ and trailing $\pt$ jet 
%for the $\mu$ + 2 jets + $\met$ final state.}
%\caption{(color online) The first two discrepant histograms in the $\mu$ + 2 jets + $\met$ exclusive final state.  
\caption{(color online) The discrepant distributions in the $\mu$ + 2 jets + $\met$ exclusive final state.  
(a) The transverse mass distribution of the $W$ boson plus second jet, (b) 
 the $\Delta {\cal R}$ between the muon and the second jet,
(c) the invariant mass distribution of the $\mu$ + second jet, and (d) $\Delta \eta$ between the highest 
$p_T$ jet and the second jet.  
``Other" contains distributions too small to list individually, $W + b \bar{b}$, diboson, 
$t \bar{t}$, and D-Y + 0 $lp$.}
\label{fig:results_shapediscrepant_mismu2j_suma}
\end{figure}

\begin{figure*}[htp]
\begin{center}
%\subfigure[] {
\label{fig:results_shapediscrepant_mismu1j_a} 
\includegraphics[width=3.0in]{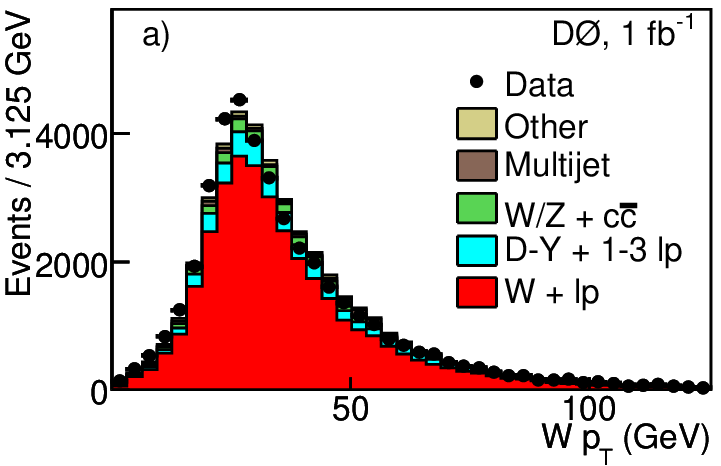}
\label{fig:results_shapediscrepant_mismu1j_b} 
\includegraphics[width=3.0in]{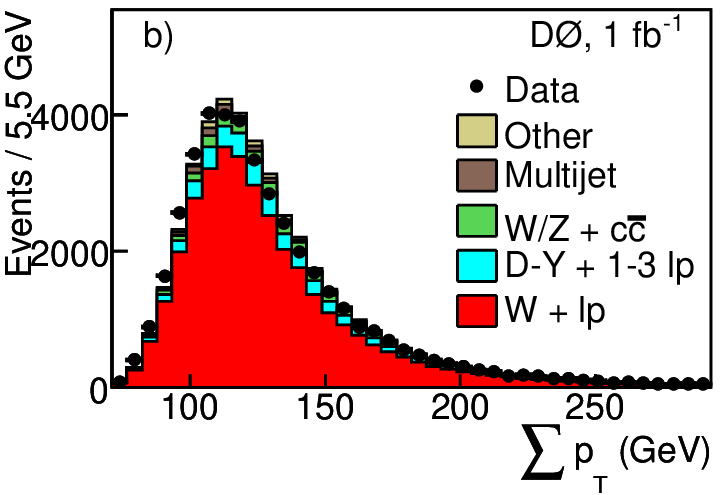}
\includegraphics[width=3.0in]{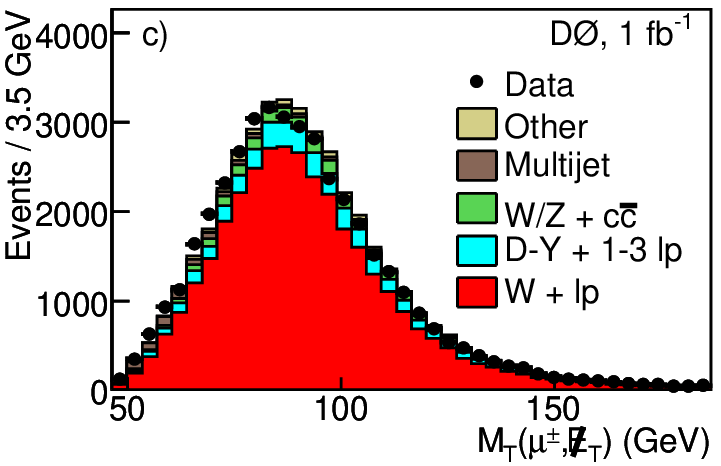}
\includegraphics[width=3.0in]{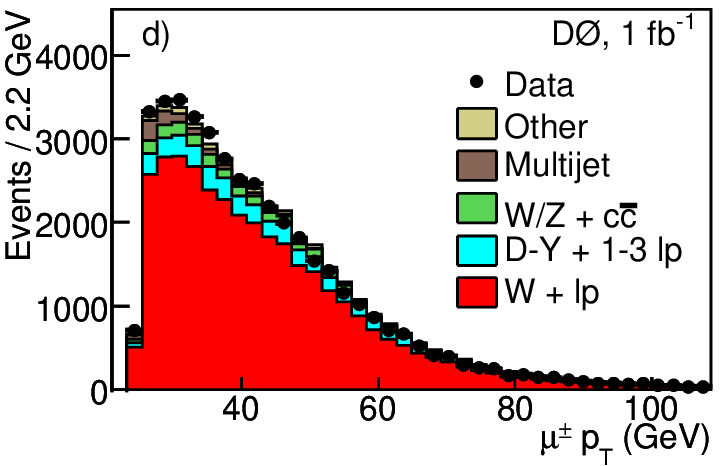}
\includegraphics[width=3.0in]{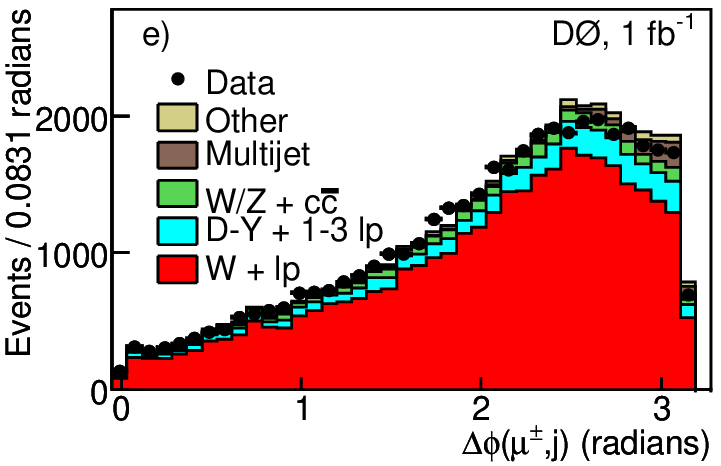}

%} 
\end{center}
%\caption{(color online) The first two discrepant histograms in the $\mu$ + 1 jet + $\met$ exclusive final state.  
\caption{(color online) The discrepant distributions in the $\mu$ + 1 jet + $\met$ exclusive final state:  
(a) the $p_T$ of the $W$ boson, (b) the sum of the scalar values of $p_T$ of the $\mu$, jet, 
and $\met$,
(c) the transverse mass of the $\mu$ and $\met$, 
(d) the $p_T$ of the $\mu$, 
and (e) 
the $\Delta \phi$ between the muon and the jet.
``Other" contains distributions too small to list individually, $W + b \bar{b}$, diboson,
$t \bar{t}$, and D-Y + 0 $lp$.
}
\label{fig:results_shapediscrepant_mismu1j_suma}
\end{figure*}

\begin{figure}[htp]
\begin{center}
%\subfigure[] {
\label{fig:results_shapediscrepant_mise2j_a} 
\includegraphics[width=3.0in]{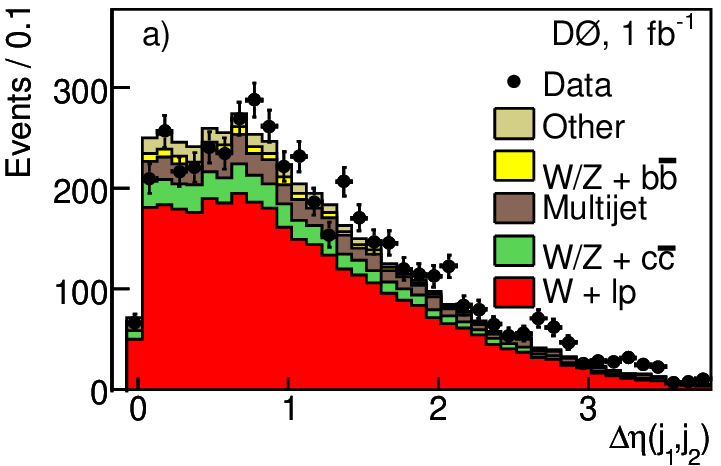}
%}

%\subfigure[] { 
\label{fig:results_shapediscrepant_mise2j_b} 
\includegraphics[width=3.0in]{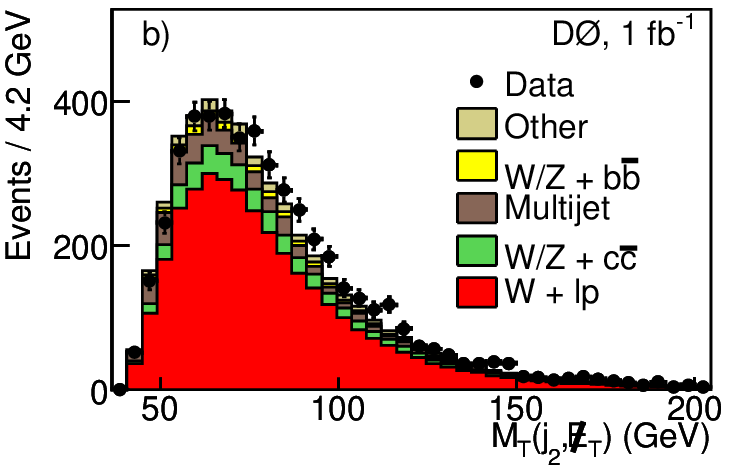}
\includegraphics[width=3.0in]{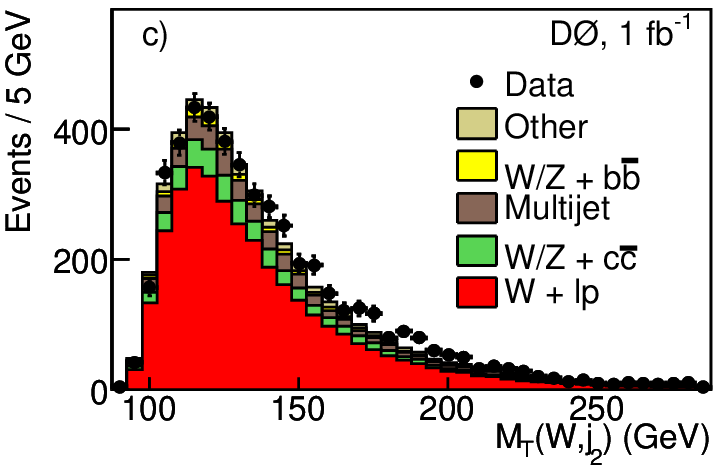}

%} 

\end{center}
\caption{(color online) The discrepant histograms in the $e$ + 2 jets + $\met$ exclusive final state.  (a) The $\Delta \eta$ between the two 
jets, (b) the transverse mass of the trailing jet and $\met$, and (c) the transverse mass distribution of the $W$ boson plus trailing jet.
``Other" contains distributions too small to list individually, diboson, D-Y, and $t \bar{t}$}
\label{fig:results_shapediscrepant_mise2j_sumb}
\end{figure}

\begin{figure}[htp]
\begin{center}
%\subfigure[] {
\label{fig:results_shapediscrepant_mise1j_a} 
\includegraphics[width=3.0in]{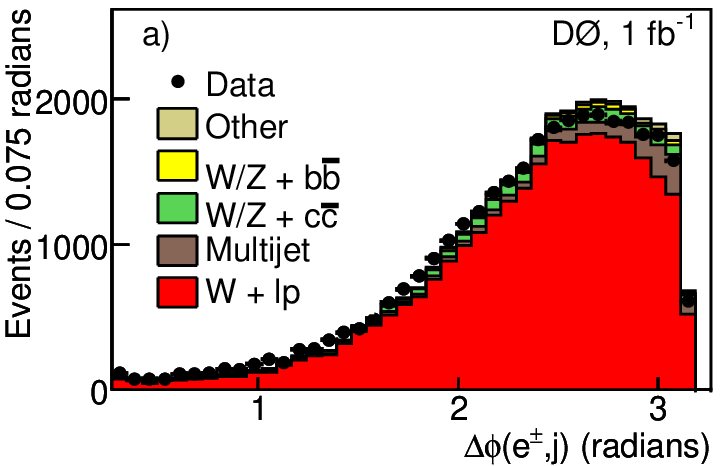}
%}

%\subfigure[] { 
\label{fig:results_shapediscrepant_mise1j_b} 
\includegraphics[width=3.0in]{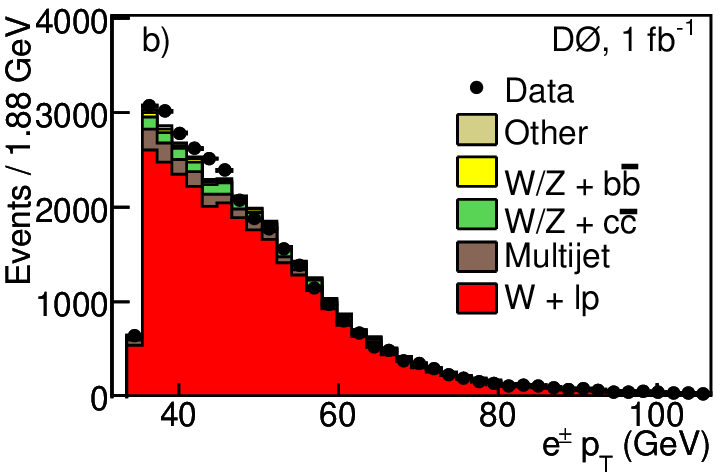}
\includegraphics[width=3.0in]{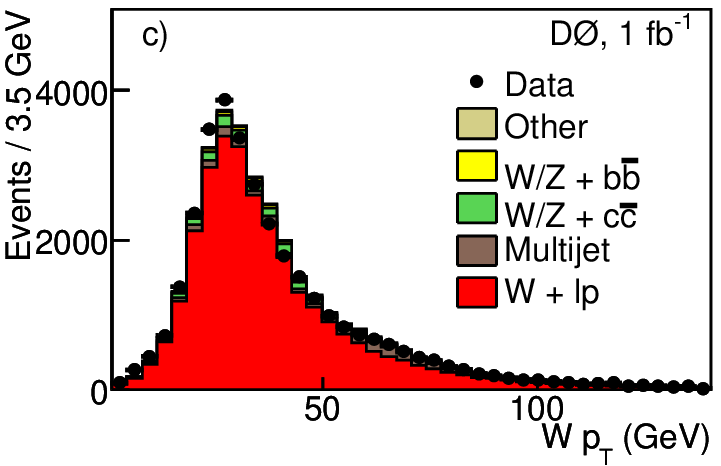}
\includegraphics[width=3.0in]{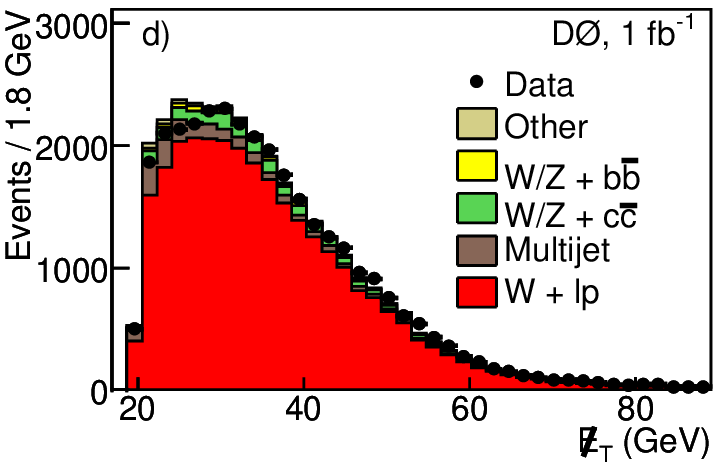}
%} 
\end{center}
\caption{(color online) The discrepant distributions in the $e$ + 1 jet + $\met$ exclusive final state.  (a)
The $\Delta \phi$ between the $e$ and $\met$, (b)
the $p_T$ of the electron, (c) the $p_T$ of the $W$ boson, and (d) the $\met$ distribution.
``Other" contains distributions too small to list individually, diboson, D-Y, and $t \bar{t}$}
\label{fig:results_shapediscrepant_mise1j_sumb}
\end{figure}

Sixteen distributions are found to be discrepant at the 3$\sigma$ level after correcting for the trials. The majority of these are related to spatial 
%distributions involving jets, low $\met$ excesses in dilepton distributions and multijet-background-dominated $\tau$ distributions. All of these types of discrepancies 
distributions involving jets. All these discrepancies 
are related to known simplifications in our modeling assumptions, e.g., no systematic uncertainties taken into account, 
aside from the adjustments made by the normalization factors.  These discrepancies would not be expected to severely 
affect the {\sc sleuth} search for new physics at high $p_T$ tails.  All 16 discrepant distributions are shown in Figs. 
\ref{fig:results_shapediscrepant_mismu2j_suma}--\ref{fig:results_shapediscrepant_mise1j_sumb} and are listed in Table \ref{tab:results_shapediscrepancies}.
In the figures, the second jet refers to the lower $p_T$ jet in the two jet final states.

%\begin{figure}[htp]
%\begin{center}

%\subfigure[] {
%} 
%\subfigure[] { 
%} 

%\end{center}
%\caption{(color online) The second two discrepant histograms in the $\mu$ + 2 jets + $\met$ exclusive final state.  
%(a) The invariant mass distribution of the $\mu$ + trailing jet, and (b) the $\Delta \eta$ between the two jets.}
%\label{fig:results_shapediscrepant_mismu2j_sumb}
%\end{figure}

%}
%\subfigure[] { 

%\begin{figure}[htp]
%\begin{center}

%\subfigure[] {
%\label{fig:results_shapediscrepant_mismu1j_c} 
%\includegraphics[width=3.0in]{used_figures/results/mu1jetmet_mupt.eps}
%}
%\subfigure[] { 
%\label{fig:results_shapediscrepant_mismu1j_d} 
%\includegraphics[width=3.0in]{used_figures/results/mu1jetmet_MT.eps}
%} 
%\end{center}
%\caption{(color online) The $\mu$ + 1 jet + $\met$ exclusive final state.  (a) The $p_T$ of the $\mu$, and (b) 
%the transverse mass of the $\mu$ and $\met$.}
%\label{fig:results_shapediscrepant_mismu1j_sumb}
%\end{figure}
%\clearpage

%\begin{figure}[htp]
%\begin{center}
%\includegraphics[width=3.0in]{used_figures/results/mu1jetmet_dphimuj.eps}
%\end{center}
%\caption{(color online) The final discrepant histogram in the $\mu$ + 1 jet + $\met$ exclusive final state, 
%the $\Delta \phi$ between the muon and the jet.}
%\label{fig:results_shapediscrepant_mismu1j_sumc}
%\end{figure}

%\begin{figure}[htp]
%\begin{center}
%\subfigure[] {
%\label{fig:results_shapediscrepant_mismu1j_e} 
%\includegraphics[width=3.0in]{used_figures/results/mu1jetmet_MTWj.eps}
%}  
%\end{center}
%\caption{(color online) Plots of the final discrepant histograms in the $\mu$ + 1 jet + $\met$ exclusive final state, the transverse mass distribution of the $W$ plus jet}
%\label{fig:results_shapediscrepant_mismu1j_e}
%\end{figure}

%\begin{figure}[htp]
%\begin{center}
%\subfigure[] {
%\label{fig:results_shapediscrepant_mise2j_c} 
%\includegraphics[width=3.0in]{used_figures/results/1e2j1pmiss_MTWj2.eps}
%} 
%\end{center}
%\caption{(color online) The final discrepant histograms in the $e$ + 2 jets + $\met$ exclusive final state, the transverse mass 
%distribution of the 
%$W$ plus trailing jet.}
%\label{fig:results_shapediscrepant_mise2j_c}
%\end{figure}

%\begin{figure}[htp]
%\begin{center}
%\subfigure[] {
%\label{fig:results_shapediscrepant_mise1j_c} 
%\includegraphics[width=3.0in]{used_figures/results/1e1j1pmiss_Wpt.eps}
%} 
%\subfigure[] {
%\label{fig:results_shapediscrepant_mise1j_d} 
%\includegraphics[width=3.0in]{used_figures/results/1e1j1pmiss_met.eps}
%} 

%\end{center}
%\caption{(color online) The final two discrepant histograms in the $e$ + 1 jets + $\met$ exclusive final state.  
% (a) The $p_T$ of the $W$, and (b) the $\met$ distribution.}
%\label{fig:results_shapediscrepant_mise1j_sumb}
%\end{figure}

%\begin{spacing}{1.0}
\ctable[
  mincapwidth = 3.0in,
%  mincapwidth = 6.0in,
  caption = {The full list of {\sc vista} results with discrepant distributions listed by final state.},
  pos = {htp},
  label = {tab:results_shapediscrepancies}
]
{ccc}
{}
{
\hline \hline
{\sc{vista}} Final State & Histogram & $\sigma$\\
%\hline
\hline
$\mu^{\pm}$ + 2 jets + $\met$ &  $M_{T}$($W$,$j_{2}$) & 4.4\\
 & $\Delta {\cal R} ( \mu, j_{2})$ & 4.4\\
 & $M$($\mu$,$j_{2}$) & 4.0\\
 & $\Delta \eta$($j_{1},j_{2}$) & 3.8 \\
\hline
{$\mu^{\pm}$ + 1 jet + $\met$} & $p_T$ ($W$) & 8.1 \\
&$ \Sigma p_T$ & 5.1 \\
& $ p_T ( \mu ) $ & 4.1 \\
&$M_{T}(\mu^{\pm},\met)$&4.1\\
&$\Delta \phi ( \mu, j)$ & 3.1 \\
\hline
$e^{\pm}$ + 2 jets +$\met$ &  $\Delta \eta$($j_{1},j_{2}$) & 4.2 \\
&  $M_{T}(j_{2}, \met)$ & 4.0 \\
& $M_T$($W$,$j_{2}$) & 3.0 \\
\hline
$e^{\pm}$ + 1 jet + $\met$  & $\Delta \phi(e^{+}, j)$ & 5.5\\
 & $p_T$($e^{\pm}$)&4.4\\
 & ${p_T}$($W$) & 3.8 \\
 & $\met$ & 3.1\\
\hline 
\hline
}
%  framerule=1.5pt,

\subsection{\label{sec:results_sleuth}{\sc sleuth}}

All {\sc vista} final states are used as input to {\sc sleuth}, and the 117 inclusive final states are folded into 31 
final states after applying global charge conjugation invariance, rebinning in the number of jets, and assuming light lepton 
universality. The two {\sc vista} final states that show broad numerical excesses are found again with the {\sc sleuth} algorithm, as expected.  No 
additional final states have a significant {\sc sleuth} output, as defined in Sec.~\ref{sec:exstates}.

\begin{figure}[htp]
%\label{fig:results_cdfcheck_a} 
\begin{center}  
\includegraphics[width=3in]{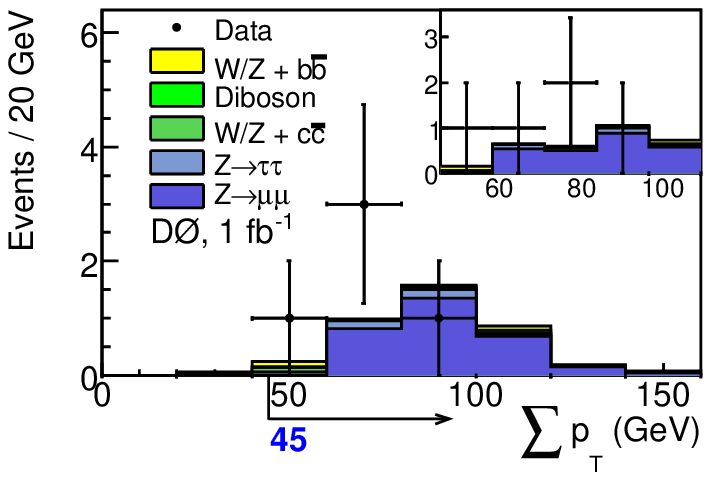}
\caption{(color online) \label{fig:results_cdfcheck_a} Check of most discrepant CDF plots from \cite{cdfRC},  $\mu^{\pm} e^{\pm}$. 
The inset shows the distribution above the $\Sigma p_{T}$ cut.
}
% The $\mathcal{P}$ values at the top right corner of the plots are the probabilities 
%before 
%final state trials 
%factors.}
\end{center}
\end{figure}

\begin{figure}[htp]
%\label{fig:results_cdfcheck_b} 
\begin{center}
\includegraphics[width=3in]{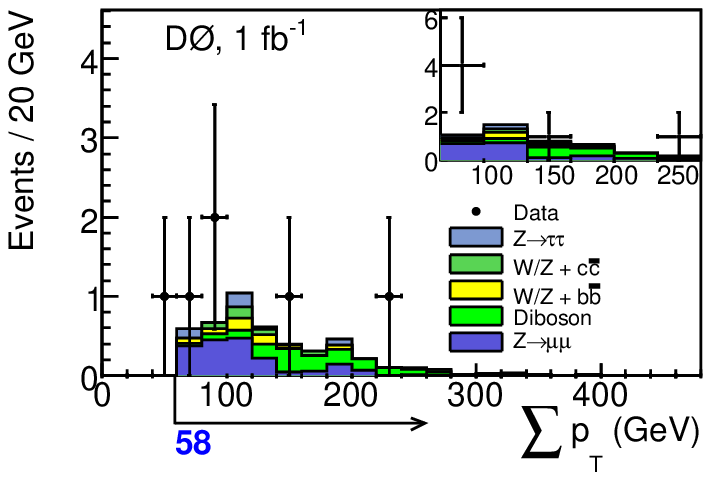} 
\label{fig:results_cdfcheck}
\end{center}
\caption{(color online)  \label{fig:results_cdfcheck_b} Check of most discrepant CDF plots from \cite{cdfRC}, $\mu^{\pm} e^{\pm}$ + $\met$. The inset shows the distribution above the $\Sigma p_{T}$ cut.
}
%The $\mathcal{P}$ values at the top right corner of the plots are the probabilities 
%before final state trials factors.}
\end{figure}

\begin{figure}[htp]
%\label{fig:results_cdfcheck_c} 
\begin{center} 
\includegraphics[width=3in]{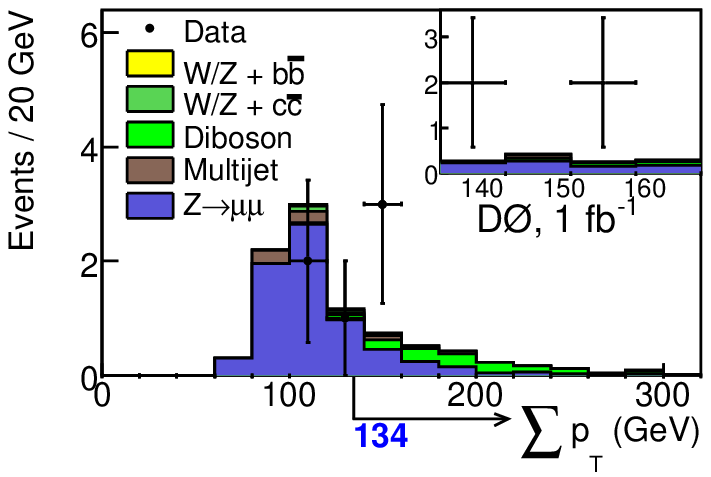}
\end{center}
\caption{(color online) \label{fig:results_cdfcheck_c} Check of most discrepant CDF plots from \cite{cdfRC}, $\ell^{\pm} \ell^{\mp} \ell^{'}$ + $\met$. The inset shows the distribution above the $\Sigma p_{T}$ cut.
}
%The $\mathcal{P}$ values at the top right corner of the plot is the probability 
%before final state 
%trials factors.}
\end{figure}

\begin{figure}[htp]
%\label{fig:results_cdfcheck_d} 
\begin{center} 
\includegraphics[width=3in]{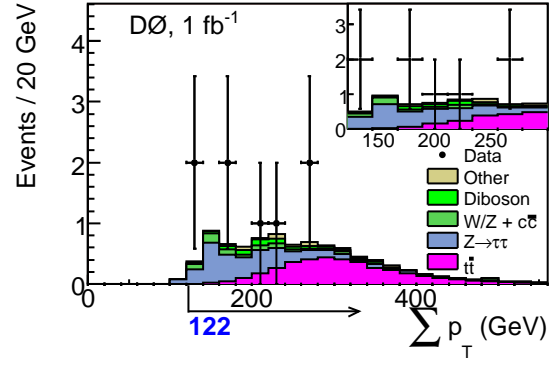}
\end{center}
\caption{(color online) \label{fig:results_cdfcheck_d} Since there are no data events in the $\mu^{\pm}$ $e^{\pm}$ + 2 jets + $\met$ final state, the distribution for  $\mu^{\pm}$ $e^{\mp}$ + 2 
jets + $\met$ 
is shown. 
%The lack of data 
%in 1 fb$^{-1}$ shows that we do not see the same data excess in that final state. 
The inset shows the distribution above the $\Sigma p_{T}$ cut.
``Other" contains the $Z \to \mu \mu$ and $W/Z + b \bar{b}$ distributions.
}
\end{figure}

In the {\sc sleuth} runs performed at CDF, using a slightly different analysis strategy, the four most interesting observed final states were  $\mu^{\pm} e^{\pm}$,  $\mu^{\pm} e^{\pm}$ + 2 jets + $\met$,  $\mu^{\pm} 
e^{\pm}$ + $\met$, and  $\ell^{\pm} \ell^{\mp}$ $\ell^{'}$ + $\met$ in $2.0$ fb$^{-1}$~\cite{cdfRC} of integrated luminosity. These states were also among the 
most discrepant observed by CDF in 0.9 pb$^{-1}$~\cite{cdfPRD} of integrated luminosity. Our 
results for these states are shown in Figs.~\ref{fig:results_cdfcheck_a}, \ref{fig:results_cdfcheck_b}, and 
\ref{fig:results_cdfcheck_c}, except for $\mu^{\pm} e^{\pm}$ + 2 jets + $\met$, for which we find no events 
with 0.16 events expected. Figure 
\ref{fig:results_cdfcheck_d} shows the similar final state, where the muon and electron are of opposite sign rather than of the same sign where CDF sees a discrepancy. %At \dzero\ with $1.07$ fb$^{-1}$, the $\mathcal{P}$ 
%value is fairly low (between 0.75 and 0.08) in Figures \ref{fig:results_cdfcheck_a} and \ref{fig:results_cdfcheck_b}, but 
None of these states are significantly discrepant in our analysis.

\begin{figure}[htp]
\begin{center}
{\includegraphics[width=3in]{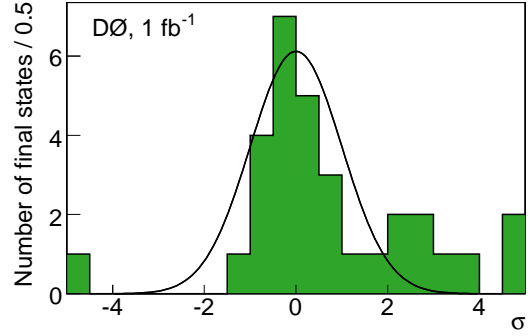}}
%{\includegraphics[height=3in,angle=270]{used_figures/results/sleuth_summary.eps}}
\caption{(color online) \label{fig:results_sleuth_summary} Distribution of final state {\sc sleuth} probabilities converted into units 
of $\sigma$ before inclusion of the final state trials factor.}
\end{center}
\end{figure}

\begin{figure}[htp]
\begin{center}
\includegraphics[width=3in]{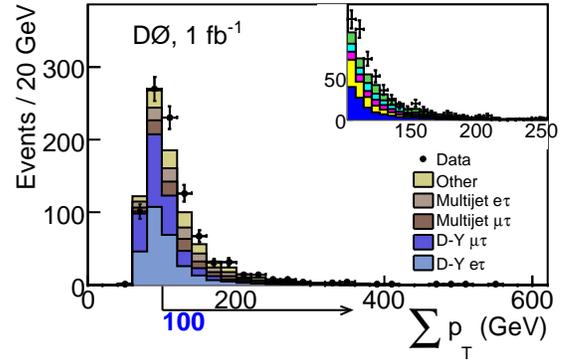}
%\label{fig:results_sleuth_etaumet}
\caption{(color online) \label{fig:results_sleuth_etaumet} {\sc sleuth} plot for $\ell^{\pm} + \tau^{\mp}$ + $\met$.
The inset shows the distribution above the $\Sigma p_{T}$ cut.  ``Other" includes D-Y $ee$ + jet events, 
D-Y $\mu\mu$ + jet events, diboson events, and $t \bar{t}$ events. }

% The $\mathcal{P}$ value at the top right corner of the plot is the probability before final state trials 
%factor.}
\end{center}
\end{figure}

%A table of the top five {\sc sleuth} final states that contain only leptons and jets is shown in Table \ref{tab:results_sleuth_states}. 
The {\sc sleuth} final states with $\tilde{\mathcal{P}} \leq 0.99$ are shown in Table \ref{tab:results_sleuth_states}. 
%An example of a distribution 
%expected to show this issue is a single lepton with $\met$, seen in Figure \ref{fig:results_sleuth_w}. 
A plot including all of the final state probabilities converted to units of $\sigma$ can be seen in 
Fig. \ref{fig:results_sleuth_summary}.
%The most discrepant final state that was not identified already in {\sc vista}, $\ell^{+}$ + $\tau^{-}$ + $\met$, is shown in Figure \ref{fig:results_sleuth_emupmiss}.
The final state $\ell^{\pm}$ + $\tau^{\mp}$ + $\met$, which was not identified as having a significant discrepancy between data and the SM expectation in 
{\sc vista}, falls close to our {\sc sleuth} threshold.  Figure~\ref{fig:results_sleuth_etaumet} shows 
the $\sum p_T$ distribution for this final state.

%In addition to the standard {\sc sleuth} algorithm, we run {\sc sleuth} with final state ordering \cite{biller}.  We create an ordered list of final states, and when calculating the trials factor for a given final state, 
%we only include final states in the count that are either higher or at the same place on the list. In this ordering we first examine the final states, listed above, in which CDF saw the largest differences first 
%~\cite{cdfRC}, with no addition trials factor from the other final states.  We then consider the remaining final states.  The most discrepant of these final states sees its $\tilde{\mathcal{P}}$ decrease from 0.96 to 
%0.32, so no additional final states have $\tilde{\mathcal{P}} < 0.001$.
%\clearpage

%Provide the table sleuth states.
%\begin{spacing}{1.0}
\ctable[
  mincapwidth = 3.0in,
%  mincapwidth = 6.0in,
  caption = {The {\sc sleuth} states with $\tilde{\mathcal{P}}<0.99$. The value of $\mathcal{P}$ represents the corresponding probability without taking into account the trial factor.},
%the $\mathcal{P}$ represents the probability that the SM background for an individual final state would have a fluctuation at 
%any cut that would be more significant than what is seen in data. The variable $\tilde{\mathcal{P}}$ is the probability that one would observe a final state with $\mathcal{P}$ less than or equal to the one 
%observed in data based on a statistical fluctuation.},
  pos = {htp},
%  framerule=1.5pt,
%  framesep=1.0pt,
  label = {tab:results_sleuth_states}
]
%{c|c|c}
{ccc}
{\tnote[a]{The value of $\tilde{\mathcal{P}}$ is not necessarily accurate below 0.001. The important check is whether the value drops below the threshold. Further discussion can be found in~\cite{cdfPRD}.}}
{
\hline
\hline
Final State & $\mathcal{P}$ & $\tilde{\mathcal{P}}$\tmark[a] \\
\hline
%$\ell^{+} \ell^{-}$ + $\met$ & $ < 3.7 \times 10^{-7}$ & $ < 2.7 \times 10^{-5}$\\
$\ell^{+} \ell^{-}$ + $\met$ & $ < 10^{-5}$ & $ < 0.001$\\
%\hline
%$\ell^{+}$ +2j + $\met$ & $ < 3.7 \times 10^{-7}$ & $ < 2.7 \times 10^{-5}$\\
$\ell^{\pm}$ +2j + $\met$ & $ < 10^{-5}$ & $ < 0.001$\\
%\hline
$\ell^{\pm}$ + $\tau^{\mp}$ + $\met$ & $ 8.9 \times 10^{-5}$ & 0.0050 \\
%\hline
$\ell^{\pm}$ + $\met$ + 1j& 0.00036 & 0.019\\
%\hline
$e^{\pm} \mu^{\mp}$ +$2b$ + $\met$ & 0.0028 & 0.12\\
%\hline
$\ell^{\pm} \tau^{\pm}$ + 2j + $\met$ & 0.0028 & 0.12\\
%\hline
$\ell^{\pm}$ + $2b$ + $\met$ & 0.0077 & 0.3\\
%\hline
$e^{\pm} \mu^{\mp}$ + $\met$ & 0.0081 & 0.31\\
%\hline
$\ell^{\pm} \tau^{\pm}$ & 0.057 & 0.91\\
%\hline
$\ell^{\pm}$ + $2b$ + 2j + $\met$ & 0.099 & 0.98 \\
%\hline
\hline
\hline
}
%\end{spacing}

%\section{Track Resolution Correction}
%Despite the additional smearing that performed for the muons and electrons, the track resolution is still found to be improperly modeled, especially for very straight tracks. The place where the effect of this is 
%most obvious is in the number of same sign electron events. The muons sign is compared in the tracker and local muon system to ensure that they match, so this is not as much of a problem with same sign muon. A 
%procedure outlined in Appendix \ref{app:track_res} is used to reweight the same sign electron and tau events to account for differences in track resolution. The other place where this mismodeling will show itself is 
%in muons with very high transverse momenta. No particular matching is used for these.

\section{\label{sec:conclusions}Conclusions}
%%%%%%%%%%%%%%%%%%%%%%%%%%%%%%%%%%%%%%%%%%%%%%%%%%%%%%%%%%%%%%%%%%%%%%%%%
% Short Title: Chapter 9
%
% Comments: Conclusions
%
%%%%%%%%%%%%%%%%%%%%%%%%%%%%%%%%%%%%%%%%%%%%%%%%%%%%%%%%%%%%%%%%%%%%%%%%
%
%
%This analysis takes a global look at D0 high-$p_T$ data to determine whether what we see is what we expect.  For this, we performed a broad search for 
%physics beyond the SM in $1.07$ fb$^{-1}$ of integrated luminosity collected in 
We have done a global study of D0 high $p_T$ data
to search for significant deviations from the standard model
expectations. This broad search for BSM physics has been done
on 1.1 fb$^{-1}$ of integrated luminosity collected in 
Run II of the Fermilab Tevatron Collider in the \dzero~ experiment. Using the {\sc vista} algorithm, a total of 117 exclusive final states and 5543 kinematic distributions were compared to the SM background 
predictions. Only two out of 117 exclusive final states, $\mu^{\pm}$ + 2 jets + $\met$ and $\mu^{+} \mu^{-} + \met$, show a statistically significant discrepancy. Given the known 
modeling difficulties in both final states together with our neglect in this study of systematical uncertainties, we cannot 
attribute the observed discrepancies to sources of physics beyond the standard model. 
A quasi-model-independent search for new physics was also performed using the algorithm {\sc sleuth} by looking for statistically significant excess at high $\sum$\pt\ in a wide array of
exclusive final states.  No additional final states cross the discovery threshold in {\sc sleuth} beyond the excesses noted by {\sc vista}.

\section{\label{sec:acknowledgements}Acknowledgements}

% acknowledgement.tex                             6 April 2011
%
We thank the staffs at Fermilab and collaborating institutions,
and acknowledge support from the
DOE and NSF (USA);
CEA and CNRS/IN2P3 (France);
FASI, Rosatom and RFBR (Russia);
CNPq, FAPERJ, FAPESP and FUNDUNESP (Brazil);
DAE and DST (India);
Colciencias (Colombia);
CONACyT (Mexico);
KRF and KOSEF (Korea);
CONICET and UBACyT (Argentina);
FOM (The Netherlands);
STFC and the Royal Society (United Kingdom);
MSMT and GACR (Czech Republic);
CRC Program and NSERC (Canada);
BMBF and DFG (Germany);
SFI (Ireland);
The Swedish Research Council (Sweden);
and
CAS and CNSF (China).
%
   % input acknowledgement

\end{document}